\documentclass[oldversion]{aa600}
\usepackage{graphicx}
\usepackage{txfonts}
\usepackage{natbib}

\newcommand{\psr}{PSR~J1911$-$5958A}
\newcommand{\gc}{NGC\,6752}
\newcommand{\kms}{km\,s$^{-1}$}
\newcommand{\ubv}{\hbox{$U\!BV$}}

\begin{document}

  \title{The masses of PSR~J1911$-$5958A and
    its white dwarf companion} 

  \author{C.\ G.\ Bassa\inst{1}
    \and M.\ H.\ van Kerkwijk\inst{2}
    \and D.\ Koester\inst{3}
    \and F.\ Verbunt\inst{1}}

  \institute{Astronomical Institute, Utrecht University, PO Box 80\,000,
    3508 TA Utrecht, The Netherlands; \email{bassa@astro.uu.nl}
    \and Dept. of Astronomy and Astrophysics, Univ. of Toronto, 60 St
    George Street, Toronto, ON M5S 3H8, Canada
    \and Institut f\"ur Theoretische Physik und Astrophysik der
    Universit\"at Kiel, 24098 Kiel, Germany
  }

  \date{Received / Accepted}

  \abstract{We present spectroscopic and photometric observations of
    the optical counterpart to \psr, a millisecond pulsar located
    towards the globular cluster \gc. We measure radial velocities
    from the spectra and determine the systemic radial velocity of the
    binary and the radial-velocity amplitude of the white-dwarf
    orbit. Combined with the pulsar orbit obtained from radio timing,
    we infer a mass ratio of
    $M_\mathrm{PSR}/M_\mathrm{WD}=7.36\pm0.25$. The spectrum of the
    counterpart is that of a hydrogen atmosphere, showing Balmer
    absorption lines up to H12, and we identify the counterpart as a
    helium-core white dwarf of spectral type DA5. Comparison of the
    spectra with hydrogen atmosphere models yield a temperature
    $T_\mathrm{eff}=10090\pm150$\,K and a surface gravity $\log
    g=6.44\pm0.20$\,cgs. Using mass-radius relations appropriate for
    low-mass helium-core white dwarfs, we infer the white-dwarf mass
    $M_\mathrm{WD}=0.18\pm0.02\,M_\odot$ and radius
    $R_\mathrm{WD}=0.043\pm0.009\,R_\odot$.  Combined with the mass
    ratio, this constrains the pulsar mass to
    $M_\mathrm{PSR}=1.40^{+0.16}_{-0.10}\,M_\odot$.  If we instead use
    the white-dwarf spectrum and the distance of \gc\ to determine the
    white-dwarf radius, we find
    $R_\mathrm{WD}=0.058\pm0.004\,R_\odot$. For the observed
    temperature, the mass-radius relations predict a white-dwarf mass
    of $M_\mathrm{WD}=0.175\pm0.010\,M_\odot$, constraining the pulsar
    mass to $M_\mathrm{PSR}=1.34\pm0.08\,M_\odot$.  We find that the
    white-dwarf radius determined from the spectrum and the systemic
    radial velocity of the binary are only marginally consistent with
    the values that are expected if \psr\ is associated with \gc. We
    discuss possible causes to explain this inconsistency, but
    conclude that our observations do not conclusively confirm nor
    disprove the assocation of the pulsar binary with the globular
    cluster.
    
  }

  \maketitle

  \section{Introduction}\label{sec:1.0}
  The equation-of-state of matter at supra-nuclear densities together
  with general relativity imply a maximum mass for a rotating neutron
  star (e.g.\ \citealt{lp04}). Conversely, a measurement of a high
  neutron-star mass constrains the equation-of-state of this matter at
  these densities.  \citet{tc99} found that neutron stars in radio
  pulsars cover only a rather narrow range in mass;
  $1.35\pm0.04\,M_\odot$. However, their sample is statistically
  dominated by mildly recycled pulsars in relativistic double
  neutron-star binaries. Considerably higher masses (up to
  $\sim\!2\,M_\odot$) are expected for millisecond pulsars with
  low-mass white-dwarf companions, since binary evolution predicts
  that several tenths of solar masses of material have been
  transferred from the progenitor of the white dwarf onto the pulsar,
  spinning it up to the currently observed (millisecond) periods (for
  reviews, see \citealt{ver93,pk94,sta04}).

  About 40 of such systems are known (see review by \citealt{kbjj05}),
  but neutron-star masses have been measured for only six of
  them. PSR~J0751+1807 contains the heaviest neutron star known to
  date and with a mass of $2.1\pm0.2\,M_\odot$ \citep{nss+05} this is
  the only system for which the mass is not consistent with a value
  near $1.4\,M_\odot$.

  With the exception of PSR~J1012+5307, these neutron-star masses are
  determinated from radio timing of the millisecond pulsar; either due
  to the detection of general-relativistic effects or due to the
  detection of secular and annual variations because of the motion of
  the Earth. For PSR~J1012+5307, the neutron-star mass was determined
  through optical spectroscopy of the white-dwarf companion to the
  pulsar. These measurements yield the radial-velocity amplitude of
  the white-dwarf orbit, which combined with the pulsar orbit,
  determines the mass ratio between the white dwarf and the pulsar. A
  model-atmosphere fit to the white-dwarf spectrum provides the
  effective temperature and surface gravity of the white
  dwarf. Combining these values with white-dwarf mass-radius relations
  yield the white-dwarf mass and radius and, through the mass ratio,
  the pulsar mass.

  In this paper, we use this method to determine the mass of the
  binary millisecond pulsar \psr. This pulsar is in a 20\,h, highly
  circular ($e<10^{-5}$) orbit around a low-mass companion and located
  at a projected offset of $6\farcm4$ from the center of the globular
  cluster \gc\ \citep{dlm+01,dpf+02}. \citet{dpf+02} argued that the
  pulsar binary is associated with the globular cluster \gc\ (cf.\ the
  discussion in Appendix~\ref{ssec:b.3}). In order to explain the
  large distance of the pulsar from the cluster center (3.3 half-mass
  radii) and the circular orbit, \citet{cpg02} investigated several
  possible scenarios. They argued that if \psr\ was ejected out of the
  core of \gc\ this may be the result of an encounter with a wide
  binary consisting of two black holes. 

  The optical counterpart to \psr\ was discovered by \citet{bvkh03}
  and confirmed by \citet{fpsd03}. It was found that the colours and
  magnitudes of the counterpart are consistent with those of a
  0.18--$0.20\,M_\odot$ helium-core white dwarf at the distance of
  \gc. The relative brightness of the counterpart ($V=22.1$) and the
  fact that the field surrounding \psr\ is not extremely crowded,
  motivated us to obtain phase-resolved spectroscopic observations of
  the companion of \psr\ and determine the mass of the pulsar. In
  principle, these observations can also be used to verify the
  membership of \psr\ with \gc\ through the systemic radial velocity
  and the white-dwarf radius, which should be consistent with values
  expected for a system associated with the globular cluster. If the
  association is confirmed, the accurate distance to the globular
  cluster provides a separate constraint on the radius of the white
  dwarf and thus its mass.

  This paper is structured as follows; in \S~\ref{sec:2.0}, we
  describe our spectroscopic observations and their reduction, as well
  as the analysis of archival photometric observations. The
  radial-velocity measurements are described in \S~\ref{sec:3.0} and
  we determine the properties of the system in \S~\ref{sec:4.0}. We
  compare our results with the work by \citet{cfpd06} and present the
  overall discussion and conclusions in \S~\ref{sec:5.0}. In
  Appendix~\ref{sec:a} we elaborate on the corrections we applied to
  the wavelength scale. Finally, we discuss the membership of \psr\
  with \gc\ in Appendix~\ref{sec:b}.

  \section{Observations and data reduction}\label{sec:2.0}
  \subsection{Spectroscopy}\label{ssec:2.1}
  Twenty-three long-slit spectra of the companion of \psr\ were
  obtained with FORS1, the Focal Reducer and Low Dispersion
  Spectrograph of the ESO VLT at Cerro Paranal, on 8 different nights
  from May to August of 2004. A summary of the observations is given
  in Table~\ref{tab:1}. Between the first and second observing run,
  the instrument was moved from Unit Telescope 1 (UT1, Antu) to UT2
  (Kueyen). The spectra were obtained with the 600 lines mm$^{-1}$
  ``B'' grism and a $1\farcs31$ slit, which gives a wavelength
  coverage from 3300--5690\,\AA. The standard-resolution collimator
  was used, resulting in a pixel size of $0\farcs2$\,pix$^{-1}$ in the
  spatial direction and 1.2\,\AA\,pix$^{-1}$ in the dispersion
  direction. All spectra had integration times of 2470\,s and were
  sandwiched between two 30\,s, $B$-band, through-the-slit images and
  preceded by one or more 30\,s $B$-band acquisition images. The
  seeing, as determined from the width of the slit profiles, varied
  between $0\farcs5$ and $1\farcs0$, with only three spectra having a
  FWHM larger than $0\farcs8$. Generally, the conditions were good,
  with photometric skies. With this setup, the spectral resolution is
  set by the seeing, which is less than the slit width in all
  observations. For our average spectra, a resolution of 4.5\,\AA\ is
  inferred from the spectra of the reference star discussed
  below. Following the FORS1 calibration plan, bias, flat-field and
  wavelength calibration frames were obtained during twilight or
  daytime afer each observing night, with the telescope pointed
  towards the zenith.

\begin{figure}
    \resizebox{\hsize}{!}{\includegraphics{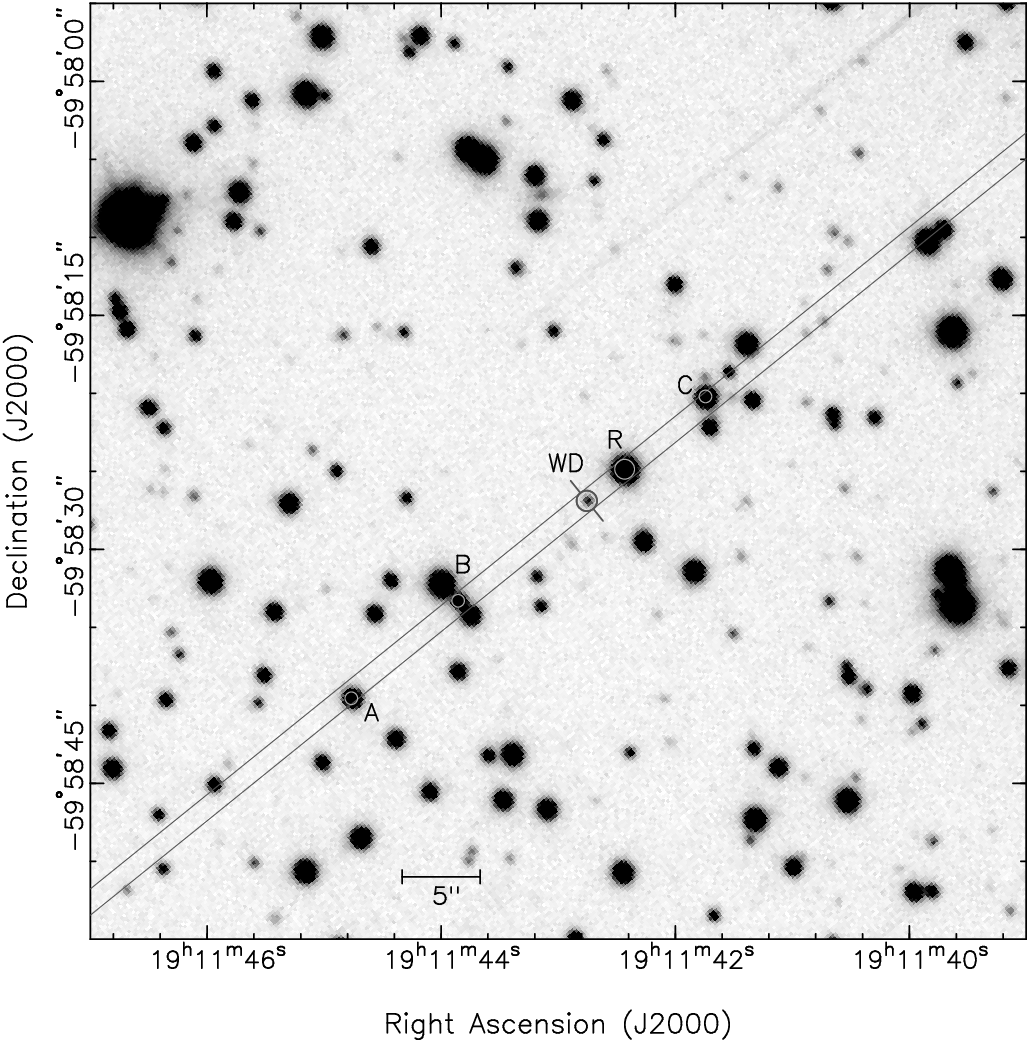}}
    \caption{The orientation of the slit on the sky. The $1\farcs31$
    slit is shown with the white-dwarf companion denoted as
    ``WD''. The nearby star used as a reference is denoted as ``R'',
    and names of some of the other stars on the slit are
    indicated. Star D is located outside the figure, to the North-West
    of \psr. This $60\arcsec\times60\arcsec$ image is an average of
    two 30\,s $B$-band acquisition images obtained during the first
    observing night.}
    \label{fig:1}
  \end{figure}

  Given the proximity of a brighter star ($V=17.3$) only $3\farcs1$ to
  the North-West of the pulsar companion, we chose to center the slit
  on both this star and the companion, see Fig.~\ref{fig:1}. We did
  this to use the star as a reference for the wavelength and flux
  calibration and to minimalize the influence of this star on the
  spectrum of the white dwarf. Besides the pulsar companion and this
  bright star (which we henceforth call the reference star or star R),
  stars A, B and C (see Fig.~\ref{fig:1}) and D also fall on the
  slit. As a result of this setup, the position angle of the slit is
  fixed on the sky and differs from the parallactic angle, by an
  amount which depends on the hour angle of the observation. Any
  effects of differential atmospheric refraction, which become
  important when one does not observe with the slit parallel to the
  parallactic angle \citep{fil82}, are largely corrected for by the
  Atmospheric Dispersion Corrector (ADC) on the FORS1 instrument.
  
  To account for slit losses and to allow for flux calibration, the
  exposure with the slit positioned closest to the parallactic angle
  (that from MJD\,53229, see Table~\ref{tab:1}) was followed by
  exposures through a $2\arcsec$ slit of both the pulsar companion
  (1600\,s) and the spectro-photometric flux standard LTT\,7987
  (30\,s; \citealt{hws+92,hsh+94}). For these the conditions were
  photometric with $0\farcs7$ seeing.

  The images were reduced with the Munich Image Data Analysis System
  (MIDAS). All images were bias-corrected with the bias values from
  the overscan regions on the FORS1 chip and flat-fielded using lamp
  exposures.  For the sky substraction we used clean regions between
  the stars along the slit. The region extended to 26\arcsec\ to each
  side of the pulsar companion, encompassing the pulsar companion, the
  reference star and star A, B and C. For star D a similar procedure
  was used. A polynomial was fitted to the spatial profile of these
  clean regions of the sky for each column in the disperion
  direction. The order of the polynomial was predominantly zero; but
  first and second order fits were used when this significantly
  increased the goodness of the fit.

  Spectra were extracted from the sky-subtracted images using an
  optimal extraction method similar to that of \citet{hor86}. Each of
  the extracted spectra was wavelength calibrated with the
  \element{He}\element{Hg}\element{Cd} wavelength calibration frames.
  Here we measured the positions of the lines in a row-averaged (in
  the spatial direction) multiplication of the wavelength calibration
  frame and the 2-dimensional (in the spatial and dispersion
  direction) slit profile of the star in question. A cubic polynomial
  fit was sufficient to describe the dispersion relation and gave
  root-mean-square (rms) residuals of less than 0.06\AA. The
  wavelength calibrations were found to be stable between different
  observing nights; the rms scatter in the zero point was 0.041\AA\
  (corresponding to 2.7\,\kms\ at 4500\AA).

  The last step was to calibrate the spectra for the instrumental
  response of the spectrograph, as derived from the observation of the
  flux standard. The spectrum of the standard was reduced in a similar
  manner as the pulsar companion and the other stars on the
  slit. Unfortunately, deriving the response was somewhat troublesome
  since the calibrated spectrum of the DA white dwarf LTT\,7987 was
  tabulated at 50\AA\ steps. With such a resolution the higher Balmer
  lines are poorly sampled which may result in systematic trends in
  the flux calibration at these wavelengths. We therefore analyzed two
  archival observations of Feige\,110 \citep{oke90}, tabulated at
  1\AA\ and 2\AA\ steps, from June 28th and December 1st, 2004. The
  observations were taken with the same grism as the pulsar companion,
  though with 5\arcsec\ MOS slits. The spectra were extracted and
  wavelength calibrated as before and corrected for atmospheric
  extinction using the average La Silla extinction curve (this
  relation is also suitable for Paranal). A comparison of the
  resulting response curves showed that these had a very similar
  shape, and that the ratio of the two curves could be well
  approximated with a linear polynomial, i.e.\ that the response was
  stable over time. We now used the response curve derived from the
  June 2004 observation of Feige\,110 and fitted it against the
  extinction-corrected response of LTT\,7987, fitting for a linear
  polynomial scaling factor. Using the $B$-band filter curve of
  \citet{bes90} and the zero point of \citet{bcp98} we obtain a
  synthetic $B$-band magnitude of 12.30 for LTT\,7987, which compares
  well to $B$=12.28 found by \citet{hws+92} and $B$=12.27 by
  \citet{lan92}.

  We corrected all spectra for atmospheric extinction and calibrated
  them using this new response. Differences in continuum flux between
  the narrow and wide slit exposures were corrected for with a scaling
  factor that depends linearly on wavelength.

  \subsection{Photometry}\label{ssec:2.2}
  We have analyzed all available FORS1 observations of the field
  containing \psr. The data consists of i) three 1500\,s $U$, five
  360\,s $B$ and eight 220\,s $V$-band images, taken with the
  high-resolution collimator (which has $0\farcs1$\,pix$^{-1}$) on 3
  different nights in 2003 March, April and May under photometric
  conditions with good seeing ($0\farcs5$--$0\farcs7$); ii) three
  32\,s $B$-band and three 13\,s $V$-band images that were obtained
  under photometric conditions with $0\farcs7$ seeing on 2003 March 31
  with the standard collimator; iii) a series of thirty 30\,s $B$-band
  acquisition images obtained prior to the spectral observations
  presented above, and iv) a series of twenty-three 600\,s $B$-band
  images obtained on 2004 August 10--15 with the high resolution
  collimator during good to moderate seeing conditions
  ($0\farcs5$--$1\farcs3$).

  \begin{figure}
    \resizebox{\hsize}{!}{\includegraphics{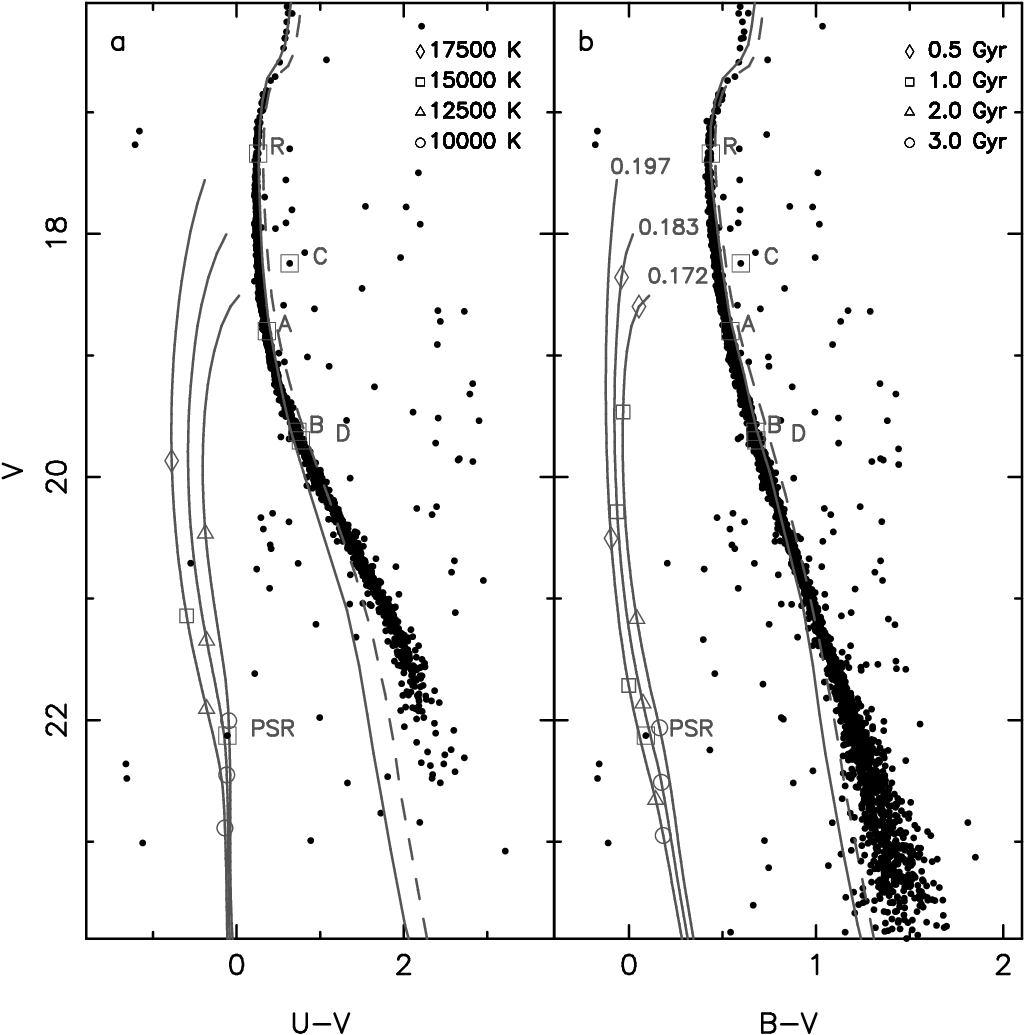}}
    \caption{Colour-magnitude diagrams of NGC\,6752, constructed from
      archival FORS1 observations. The stars located on the slit are
      labelled. Shown as solid lines to the left of the cluster
      main-sequence are three $Z=0.0010$ helium-core white dwarf
      cooling models of \citet{sarb02}. The masses of these models are
      as shown (in $M_\odot$), and temperatures are indicated along
      the track in panel~\textbf{a}, while cooling ages are shown in
      panel~\textbf{b}. Also shown are two isochrones from \citet{gbbc00}
      for an age of 14.1\,Gyr and a metallicity of $Z=0.0004$ (solid
      line) and one for an age of 12.6\,Gyr with $Z=0.0010$ (dashed
      line). All models are placed at a distance of $(m-M)_V=13.24$
      with a reddening of $E_{B-V}=0.040$, as determined by
      \citet{gbc+03}.}
    \label{fig:2}
  \end{figure}
   
  \begin{figure}
    \resizebox{\hsize}{!}{\includegraphics{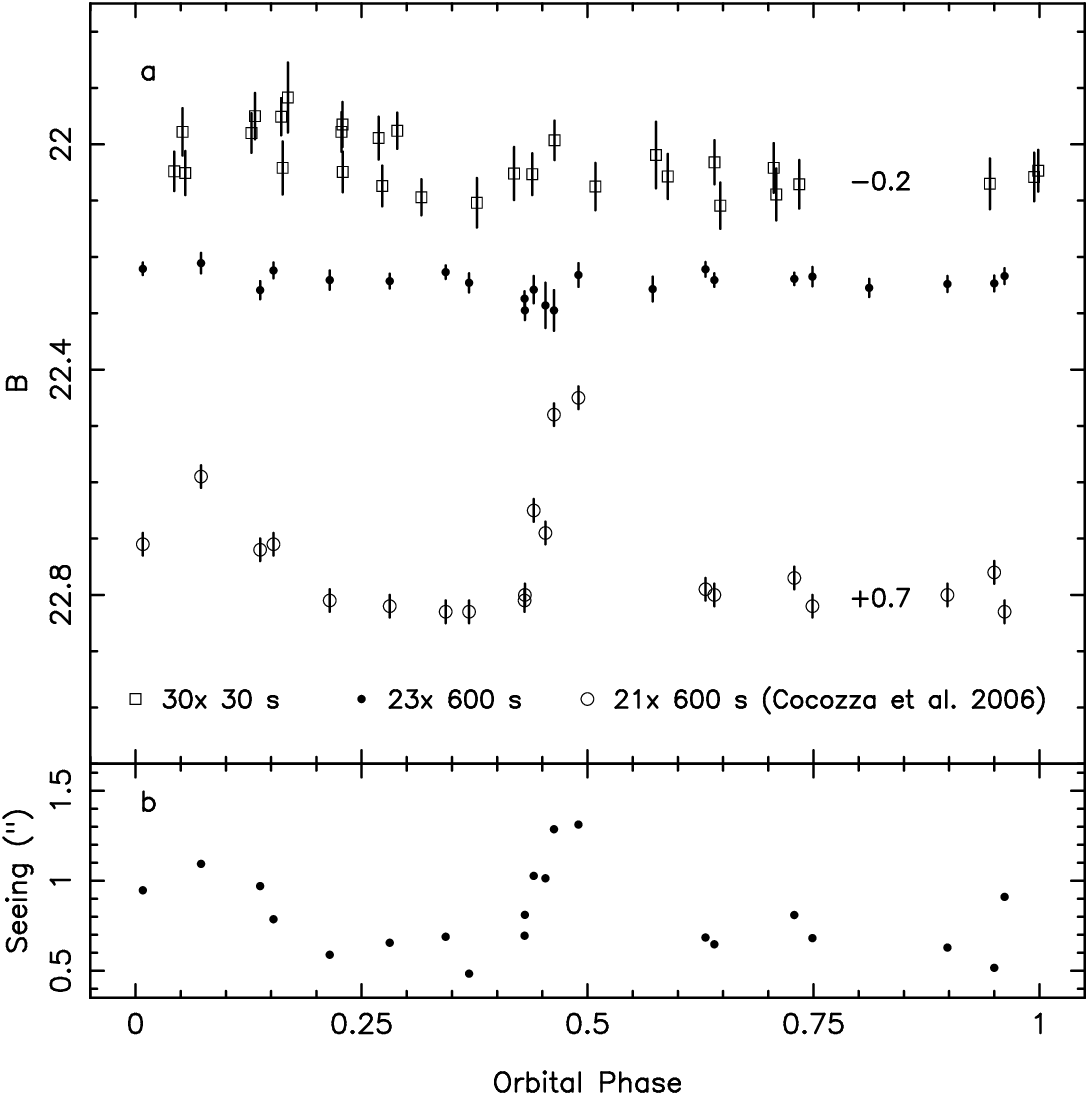}}
    \caption{\textbf{a} The $B$-band magnitude of the white-dwarf
    companion of \psr\ as a function of orbital phase for the
    magnitudes determined from the thirty 30\,s exposures (\emph{open
    squares}, offset by $-0.2$\,mag) and the twenty-three 600\,s
    $B$-band exposures (\emph{black dots}). Also shown are the
    $B$-band magnitudes determined by \citet{cfpd06} from the same
    600\,s exposures (\emph{open circles}, read off from their
    Fig.\,3, but offset by +0.7\,mag).  \textbf{b} The seeing of the
    600\,s $B$-band exposures, determined from the width of the
    point-spread-function. Note that the variations in the seeing and
    the magnitudes as measured by \citet{cfpd06} are highly
    correlated, indicating that their meaurements are corrupted (see
    \S~\ref{sec:5.0} for details).}
    \label{fig:3}
  \end{figure}

  All images were bias-subtracted and flatfielded using twilight
  flats. The DAOPHOT II package (\citealt{ste87}), running inside
  MIDAS, was used for the photometry on the averaged images. We
  followed the recommendations of \citet{ste87}, obtaining
  instrumental magnitudes through point-spread function (PSF)
  fitting. The $B$ and $V$-band observations of March 31, 2003 include
  20 photometric standards from \citet{ste00} of which 12 were
  unsaturated. The instrumental PSF magnitudes of these stars were
  directly compared against the calibrated values to derive zero
  points and colour terms (no extinction coefficients are needed since
  the standards and target are at the same airmass), giving rms
  residuals of 0.01\,mag in $B$ and 0.03\,mag in $V$. For the
  calibration of the $U$-band observations, we used 4 standard stars
  from the PG\,1657+078 field \citep{lan92}. We fitted for zero point
  and colour term, using the standard ESO extinction coefficients of
  0.46 and 0.25\,mag per airmass for $U$ and $B$-band
  respectively. These calibrations have rms residuals of 0.02\,mag in
  $U$ and $B$. The difference between the two $B$-band calibrations
  was less than 0.01\,mag, hence we expect our zero-point
  uncertainties in $B$ to be less than 0.02\,mag, less than 0.03\,mag
  in $V$, and allowing for the uncertainty in the $U$-band extinction
  coefficient, less than 0.05\,mag in $U$. The resulting magnitudes of
  the stars on the slit are tabulated in Table~\ref{tab:2}.

  Colour-magnitude diagrams were constructed from the photometry and
  are shown in Fig.~\ref{fig:2}. We find that star R has the magnitude
  and colours of a cluster turn-off star while stars A, B and D are
  located further down the cluster main sequence. Star C, on the other
  hand, is not located on the main sequence, and is about a magnitude
  brighter than cluster stars with the same $U-V$ and $B-V$
  colours. The pulsar companion is blue with respect to the cluster
  main-sequence by about 2\,mag in $U-V$ and more than 1\,mag in
  $B-V$.

  We checked for variability of the pulsar companion using the
  twenty-three 600\,s $B$-band images and the thirty 30\,s $B$-band
  acquisition images. With a pixel scale of $0\farcs1$\,pix$^{-1}$,
  the 600\,s images are severely oversampled, and we rebinned the
  images to a pixel scale of $0\farcs2$\,pix$^{-1}$ (averaging every
  $2\times2$ pixels). Next, instrumental magnitudes were determined
  through PSF fitting and calibrated to the photometry presented
  above. The resulting magnitudes are shown in Fig.~\ref{fig:3}. In
  both the 600\,s and the 30\,s images, the magnitudes of the pulsar
  companion do not significantly vary with orbital phase. For example,
  the rms scatter for the white dwarf around the average value in the
  600\,s and 30\,s images is only 0.02\,mag and 0.05\,mag,
  respectively, and these values are comparable to that of stars of
  similar brightness.

  \begin{table}
    \caption[]{VLT/FORS1 photometry of the white-dwarf companion of
      \psr\ (denoted with WD) and stars on the slit. The nomenclature
      of the stars is according to Fig.~\ref{fig:1}, while star D is
      located outside the figure, on the North-West side of the
      slit. The uncertainties listed in parentheses are instrumental,
      i.e., they do not include the zero-point uncertainty in the
      photometric calibration (0.05 mag in $U$, 0.02 mag $B$ and 0.03
      mag in $V$). The celestial positions were obtained using the
      procedure outlined in \citet{bvkh03}.}
    \label{tab:2}
    \begin{tabular}
      {l@{\hspace{0.15cm}}
	l@{\hspace{0.15cm}}
	l@{\hspace{0.15cm}}
	l@{\hspace{0.15cm}}
	l@{\hspace{0.15cm}}
	l@{\hspace{0.15cm}}
      }
      \hline
      \hline
      ID & \multicolumn{1}{c}{$\alpha_\mathrm{2000}$} & 
      \multicolumn{1}{c}{$\delta_\mathrm{2000}$} & 
      \multicolumn{1}{c}{$U$\phantom{0}} & 
      \multicolumn{1}{c}{$B$\phantom{0}} &
      \multicolumn{1}{c}{$V$\phantom{0}} \\
      & $\phantom{00}^\mathrm{h}\phantom{00}^\mathrm{m}\phantom{00}^\mathrm{s}$
      &$\phantom{00}\degr\phantom{00}\arcmin\phantom{00}\arcsec$ & & & \\
      \hline
      WD & 19 11 42.753 & -59 58 26.89 & 22.02(5) & 22.22(3) & 22.13(2) \\[0.5mm]
      R  & 19 11 42.432 & -59 58 24.90 & 17.60(5) & 17.78(1) & 17.34(1) \\
      A  & 19 11 44.768 & -59 58 39.58 & 19.16(5) & 19.34(1) & 18.80(1) \\
      B  & 19 11 42.854 & -59 58 33.34 & 20.36(5) & 20.31(1) & 19.63(1) \\
      C  & 19 11 41.742 & -59 58 20.25 & 18.88(5) & 18.84(1) & 18.24(1) \\
      D  & 19 11 35.911 & -59 57 45.35 & 20.47(5) & 20.38(1) & 19.70(1) \\
      \hline
    \end{tabular}
  \end{table}

  \begin{table*}
    \begin{minipage}[t]{\textwidth}
      \centering
      \caption[]{Radial-velocity measurements of the white-dwarf
      companion of \psr\ and four stars on the slit. To put these
      velocities on an absolute scale, a velocity offset of
      $-39\pm3$\,\kms\ should be added (see
      Appendix\,\ref{sec:a}).}\label{tab:1}
      \renewcommand{\footnoterule}{}
      \begin{tabular}{
	  c@{\hspace{0.5cm}}
	  c@{\hspace{0.6cm}}
	  r@{\hspace{0.5cm}}
	  r@{\hspace{0.5cm}}
	  r@{\hspace{0.5cm}}
	  r@{\hspace{0.5cm}}
	  r@{\hspace{0.5cm}}
	}
	\hline\hline
	& & \multicolumn{1}{c}{$v_\mathrm{WD}$} &
	\multicolumn{1}{c}{$v_\mathrm{R}$} &
	\multicolumn{1}{c}{$v_\mathrm{A}$} &
	\multicolumn{1}{c}{$v_\mathrm{C}$} &
	\multicolumn{1}{c}{$v_\mathrm{D}$} \\
	MJD$_\mathrm{bar}$\footnote{The time of the observation at
	  mid-exposure, corrected to the solar system barycenter.} &
	  $\phi_\mathrm{b}$\footnote{Using the ephemeris of
	  \citet{dpf+02}: $T_\mathrm{asc} =
	\mathrm{MJD}\, 51919.2064780(3)$,
	$P_\mathrm{b}=0.837113476(1)$\,days.} &
	\multicolumn{1}{c}{(\kms)} & \multicolumn{1}{c}{(\kms)} &
	\multicolumn{1}{c}{(\kms)} &
	  \multicolumn{1}{c}{(\kms)} & \multicolumn{1}{c}{(\kms)} \\
	\hline

	53147.38143 & 0.1590 &  $-$29 $\pm$ 22 &     12.9 $\pm$ 8.1 &     18 $\pm$ 11 &     34 $\pm$ 16 &     21 $\pm$ 20 \\
	53198.11569 & 0.7664 &      3 $\pm$ 29 &      6.8 $\pm$ 7.8 &   $-$1 $\pm$ 11 &     18 $\pm$ 16 &     11 $\pm$ 22 \\
	53204.15036 & 0.9752 & $-$233 $\pm$ 26 &     11.5 $\pm$ 8.2 &     23 $\pm$ 11 &     19 $\pm$ 16 &   $-$1 $\pm$ 20 \\
	53204.19242 & 0.0254 & $-$219 $\pm$ 22 &     17.8 $\pm$ 8.1 &     13 $\pm$ 12 &     45 $\pm$ 16 &     13 $\pm$ 21 \\
	53204.22978 & 0.0701 & $-$160 $\pm$ 25 &      8.8 $\pm$ 8.5 &     17 $\pm$ 11 &     58 $\pm$ 15 &     12 $\pm$ 21 \\
	53206.21868 & 0.4459 &    218 $\pm$ 23 &     16.8 $\pm$ 8.6 &      7 $\pm$ 11 &     33 $\pm$ 15 &      6 $\pm$ 22 \\
	53206.25549 & 0.4899 &    262 $\pm$ 20 &     10.8 $\pm$ 8.1 &     12 $\pm$ 11 &     38 $\pm$ 15 &     23 $\pm$ 21 \\
	53206.29500 & 0.5371 &    245 $\pm$ 23 &     10.3 $\pm$ 8.5 &      3 $\pm$ 11 &     31 $\pm$ 16 &     10 $\pm$ 20 \\
	53210.19705 & 0.1983 &      6 $\pm$ 25 &     15.4 $\pm$ 8.8 &      2 $\pm$ 11 &     35 $\pm$ 16 &     20 $\pm$ 20 \\
	53210.23098 & 0.2388 &     12 $\pm$ 22 &   $-$0.3 $\pm$ 8.0 &   $-$1 $\pm$ 12 &     30 $\pm$ 15 &      3 $\pm$ 20 \\
	53210.27816 & 0.2952 &     77 $\pm$ 24 &   $-$6.4 $\pm$ 7.6 &     12 $\pm$ 11 &     38 $\pm$ 16 &      2 $\pm$ 21 \\
	53210.31250 & 0.3362 &    148 $\pm$ 27 &      3.6 $\pm$ 7.8 &   $-$1 $\pm$ 11 &     33 $\pm$ 16 &     23 $\pm$ 26 \\
	53229.01322 & 0.6748 &    116 $\pm$ 23 &      6.2 $\pm$ 8.1 &      8 $\pm$ 11 &     27 $\pm$ 16 &   $-$1 $\pm$ 22 \\
	53231.02712 & 0.0805 & $-$186 $\pm$ 25 &      0.3 $\pm$ 8.1 &      8 $\pm$ 11 &     12 $\pm$ 16 &  $-$12 $\pm$ 22 \\
	53231.06463 & 0.1253 & $-$179 $\pm$ 27 &      3.8 $\pm$ 8.2 &      6 $\pm$ 11 &     19 $\pm$ 16 &   $-$8 $\pm$ 20 \\
	53231.11869 & 0.1899 &  $-$64 $\pm$ 26 &     16.3 $\pm$ 7.8 &      6 $\pm$ 11 &     33 $\pm$ 16 &     22 $\pm$ 21 \\
	53231.17422 & 0.2562 &     35 $\pm$ 25 &   $-$7.0 $\pm$ 7.9 &      4 $\pm$ 11 &     22 $\pm$ 16 &   $-$6 $\pm$ 20 \\
	53231.22574 & 0.3177 &    153 $\pm$ 29 &      7.7 $\pm$ 8.2 &   $-$6 $\pm$ 10 &     34 $\pm$ 16 &   $-$6 $\pm$ 21 \\
	53232.00988 & 0.2544 &     57 $\pm$ 22 &      1.2 $\pm$ 7.7 &      5 $\pm$ 11 &      9 $\pm$ 16 &   $-$8 $\pm$ 19 \\
	53232.04617 & 0.2978 &     42 $\pm$ 24 &     11.4 $\pm$ 8.1 &     13 $\pm$ 11 &     21 $\pm$ 15 &      9 $\pm$ 20 \\
	53232.08482 & 0.3439 &    142 $\pm$ 21 &     23.9 $\pm$ 8.2 &     15 $\pm$ 10 &     38 $\pm$ 15 &     16 $\pm$ 20 \\
	53232.13559 & 0.4046 &    200 $\pm$ 22 &     12.7 $\pm$ 8.4 &      0 $\pm$ 11 &     37 $\pm$ 17 &   $-$5 $\pm$ 20 \\
	53232.18610 & 0.4649 &    273 $\pm$ 32 &   $-$0.6 $\pm$ 9.2 &      3 $\pm$ 10 &     32 $\pm$ 17 &   $-$1 $\pm$ 21 \\
	\hline
      \end{tabular}
    \end{minipage}
  \end{table*}

  \section{Radial velocities}\label{sec:3.0}
  Radial velocities of the companion and the five stars on the slit
  were determined by comparing them with template spectra. In case of
  the white dwarf this was done iteratively, where we used the
  hydrogen atmosphere models from which we determined the surface
  gravity and effective temperature (see \S~\ref{ssec:4.2} and
  Fig.\,\ref{fig:5}) as a template. A best-fit model was first
  determined for one of the single spectra. This model was then used
  as a velocity template to measure the velocities of the other
  spectra. These spectra were shifted to zero velocity and
  averaged. The final velocity template was found by fitting a new
  atmosphere model against the averaged spectrum. The actual
  velocities were measured by minimizing a $\chi^2$ merit function, as
  defined in \citet{bkk03}, fitting for velocity and a 2nd order
  polynomial modelling continuum differences.

  In the case of the reference star we used a somewhat different
  approach. Here, a template was constructed from fitting Lorentzian
  profiles to a single, normalized spectrum of the reference
  star. Eight lines (\element{H}$\beta$ up to \element{H}11 and
  \element{Ca}~K, but without the blend of \element{H}$\epsilon$ and
  \element{Ca}~H) were simultaneously fitted, fitting for depth and
  width, but forcing the velocity to be the same for all lines. The
  resulting template was shifted to zero velocity and fitted against
  normalized spectra of the reference star by again minimizing a
  $\chi^2$ merit function. To test for the stability of the radial
  velocities, we also used this template to determine radial
  velocities of the other four stars on the slit. These stars also
  display the hydrogen Balmer lines and \element{Ca}~K, though the
  lines are not as strong as those of the reference star (see
  Fig.~\ref{fig:a2}).

  \begin{figure}
    \resizebox{\hsize}{!}{\includegraphics{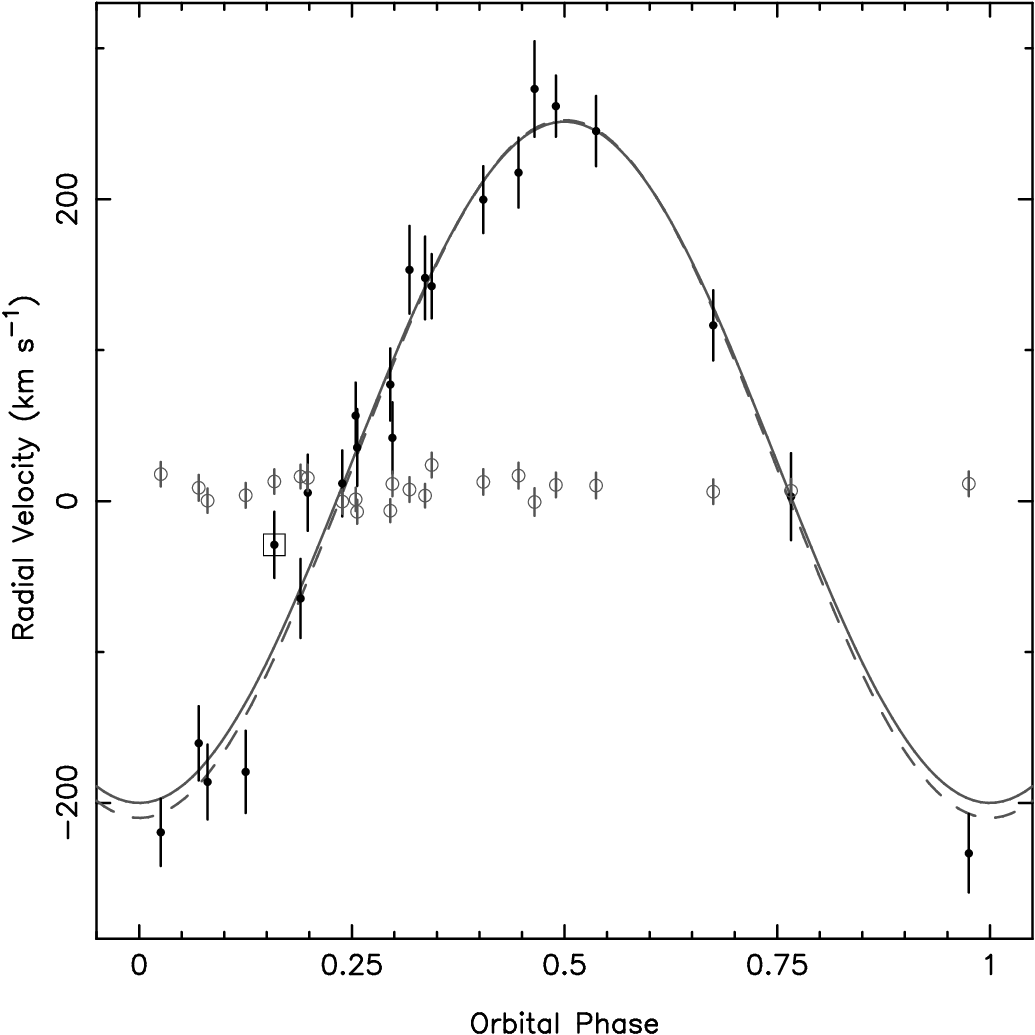}}
    \caption{The radial velocities of the white-dwarf companion of
      \psr\ (\emph{black dots}) and of the reference star~R
      (\emph{open circles}). The solid line represents the best-fit
      model for a circular orbit, using all data points, with the time
      of the ascending node passage and the orbital period fixed to
      the values determined from the radio-timing solution by
      \citet{dpf+02}. The dashed line represents the best-fit model
      excluding the boxed data point. To put these velocities on an
      absolute scale, a velocity offset of $-39\pm3$\,\kms\ should be
      added (see Appendix\,\ref{sec:a}).}
    \label{fig:4}
  \end{figure}

  The radial velocity of the pulsar companion varied by as much as
  470\,\kms\ between the different spectra, fully within the
  expectations for this system. From radio timing, it was found that
  the eccentricity of the orbit is $e<10^{-5}$ \citep{dpf+02}. Hence,
  we fitted the barycentric radial velocities of the pulsar companion
  to a circular orbit, with the orbital period and time of the
  ascending passage node fixed to the values determined from the
  radio-timing solution by \citet{dpf+02}. We find a radial-velocity
  amplitude $K_\mathrm{WD}=215\pm10$\,\kms\ and a systemic velocity
  $\gamma=-30\pm7$\,\kms\ for a reduced chi-squared $\chi^2_\nu=1.67$
  with 21 degrees-of-freedom.  Here, the errors on $K_\mathrm{WD}$ and
  $\gamma$ are scaled to give $\chi^2_\nu=1$.

  The velocity of the reference star, however, varied over a
  120\,\kms\ range, much larger than the 7 to 10\,\kms\ error on the
  individual velocities. We feared that these velocity variations
  might also be a result of binarity, but the other stars on the slit
  displayed similar variations in velocity. In particular, the
  velocities of these stars displayed a trend when compared against
  the local hour angle of the observation, where the velocity
  decreased by about 16\,\kms\ per hour prior or after
  culmination. Upon closer investigation this trend was found to be
  caused by two separate, systematic, effects. Because the effects are
  systematic, they can be corrected for.

  For the first correction, we applied a shift to the wavelength
  calibration of each spectrum based on the difference between the
  measured wavelength of the \ion{O}{i} $\lambda5577$ night sky
  emission line in the spectrum and the labaratory value. This
  correction removes the global decrease of the velocities as a
  function of hour angle. The remaining scatter in the radial
  velocities is largely removed by correcting for the second effect,
  which is due to errors in the centering of the stars on the
  slit. For this correction, we determined the position of star R with
  respect to the center of the slit in the through-the-slit images
  obtained before and after each spectrum. We apply this offset as a
  shift in wavelength to the wavelength calibration of the
  corresponding spectrum. In Appendix\,\ref{sec:a} we describe these
  corrections in detail, while Table\,\ref{tab:2} lists the velocities
  that were determined from the corrected wavelength calibrations.

  We now use the corrected radial velocities to determine the
  radial-velocity orbit of the white-dwarf companion of \psr. Again
  fitting for a circular orbit, we find a radial-velocity amplitude
  $K_\mathrm{WD}=226\pm9$\,\kms\ and a systemic velocity
  $\gamma=26\pm6$\,\kms\ ($\chi^2_\nu=1.45$ for 21
  degrees-of-freedom). The errors on $K_\mathrm{WD}$ and $\gamma$ are
  again scaled to give $\chi^2_\nu=1$. This fit is represented in
  Fig.~\ref{fig:4} with the solid line. If we exclude the single point
  (the boxed point in Fig.~\ref{fig:4}) that lies $3.2\sigma$ away
  from the best-fit, the fit improves to $\chi^2_\nu=1.00$, giving
  $K_\mathrm{WD}=231\pm8$\,\kms\ and $\gamma=21\pm5$\,\kms, depicted
  by the dashed curve in Fig.~\ref{fig:4}. Remarkably, this outlier
  corresponds to the spectrum taken during the first observing run,
  when, as mentioned in \S~\ref{sec:2.0}, FORS1 was still on UT1,
  unlike all other spectroscopic observations, when it was on UT2. We
  do not understand, however, how this could cause a difference, since
  differences in the flat-fields, wavelength calibration, and flux
  calibration should all be corrected for.

  If we fit a circular orbit against the radial velocities of the
  pulsar companion relative to the radial velocities of the reference
  star R, we obtain $K_{\Delta v}=225\pm10$\,kms, $\gamma_{\Delta
  v}=17\pm6$\,\kms, $\chi^2_\nu=1.43$ for 21 degrees-of-freedom. Again
  excluding the outlier gives $K_{\Delta v}=231\pm9$\,\kms,
  $\gamma_{\Delta v}=13\pm6$\,\kms, with $\chi^2_\nu=1.10$ for 20
  degrees-of-freedom.

  The differences in the radial velocity amplitudes $K_\mathrm{WD}$
  are consistent within the errors. The same holds for the systemic
  velocities $\gamma$ measured from the absolute and the relative
  velocities. For the remainder of this paper we will use the fit
  using absolute velocities, without the $3.2\sigma$ outlier;
  $K_\mathrm{WD}=231\pm8$\,\kms\ and $\gamma=-18\pm6$\,\kms\ (here we
  corrected the systematic velocity for the $-39\pm3$\,\kms\ velocity
  offset which we determined in Appendix\,\ref{sec:a}).

  \section{System properties}\label{sec:4.0}
  We use our measurements to determine the properties of the white
  dwarf and the pulsar. In our analysis, we distinguish between
  results that do and that do not depend on the assumption that the
  pulsar is a member of \gc. We will see that our conclusions depend
  on that assumption; we will address this in detail in
  Appendix~\ref{sec:b}.

  \subsection{Minimum white dwarf mass}\label{ssec:4.1}
  The radio timing observations by \citet{dpf+02} yielded a projected
  semi-major axis of the pulsar orbit of $a_\mathrm{PSR} \sin
  i/c=1.206045\pm0.000002$\,s, which, together with the orbital period
  $P_\mathrm{b}$ implies a radial-velocity amplitude of
  $K_\mathrm{PSR}=31.40986\pm0.00005$\,\kms. Combining this with the
  radial-velocity amplitude of the white dwarf determines the mass
  ratio $q=M_\mathrm{PSR}/M_\mathrm{WD} = K_\mathrm{WD}/K_\mathrm{PSR}
  = 7.36\pm0.25$. Here, the error on $q$ is dominated by the
  uncertainty in $K_\mathrm{WD}$.

  We can use the mass ratio and the constraint that the inclination
  must be less than or equal to $90\degr$ to determine a lower limit
  to the white-dwarf mass. For this, we use the pulsar mass-function
  $f(M_\mathrm{PSR})=M_\mathrm{WD}^3 \sin^3
  i/(M_\mathrm{WD}+M_\mathrm{PSR})^2=(2.687603\pm0.000013)\times10^{-3}\,M_\odot$,
  so that we can write $M_\mathrm{WD} \sin^3 i=(1+q)^2
  f(M_\mathrm{PSR})$. Setting the inclination at $i=90\degr$ and using
  the mass ratio $q$ as determined above we find a $1\sigma$ lower
  limit of $M_\mathrm{WD}>0.177\,M_\odot$. The $2\sigma$ lower limit
  is $M_\mathrm{WD}>0.166\,M_\odot$.

  \subsection{Effective temperature and surface gravity}\label{ssec:4.2}
  The atmospheric parameters for the white dwarf were determined by
  fitting theoretical model atmospheres to the average of the
  velocity-corrected spectra (see \S~\ref{sec:3.0}). The theoretical
  models were taken from a grid of pure hydrogen models usually
  applied to normal DA white dwarfs, but extending down to surface
  gravities of $\log g = 5$. The methods and input physics are
  described in more detail in \citet{fkb97} and \citet{hkh+98}. The
  best-fitting parameters are found with a Levenberg-Marquardt type
  $\chi^2$ algorithm \citep{ptvf92}. We also use a second, newly
  developed algorithm, which is less sophisticated but more
  transparent and robust than the Levenberg-Marquardt method. In
  essence it determines the $\chi^2$ values for the models of the grid
  around the minimum and then fits the $\chi^2$ surface with a
  paraboloid, from which the parameters and errors corresponding to
  the minimal $\chi^2$ can be calculated. This avoids excessive
  interpolations between the models in the grid, which sometimes leads
  to artificial small-scale structure of the $\chi^2$ surface.  The
  results between the two methods did not differ significantly; the
  values given below are from the second method.

\begin{figure*}
    \centering
    \includegraphics[width=17cm]{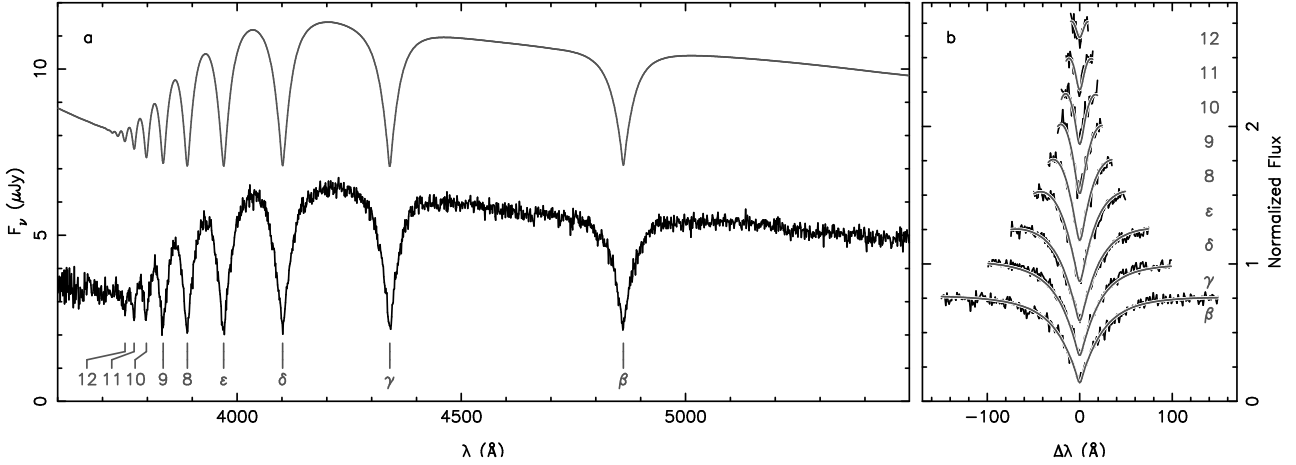}
    \caption{Spectrum of the white-dwarf companion to \psr. The lower
    curve in panel~\textbf{a} shows the average of the 23 individual
    spectra, shifted to zero velocity. The hydrogen Balmer lines are
    seen from H$\beta$ up to H12 as indicated. The top curve is the
    best-fit model spectrum, with $T_\mathrm{eff}=10090$\,K and $\log
    g=6.44$\,cgs. It is shifted upwards by
    5$\mu$Jy. Panel~\textbf{b} shows the flux-normalized line profiles
    superposed on the modelled profiles. The H$\beta$ profile is
    shifted a quarter unit downwards from unity, while those of
    H$\delta$ up to H12 are shifted upwards by multiples of the same
    amount.}
    \label{fig:5}
  \end{figure*}

  For the fit we used the spectral ranges from 3740--4440 and
  4760--5030\,\AA, which contain the Balmer lines. The model was
  fitted to the observed spectrum and the $\chi^2$ calculated from the
  fit to the continuum-normalized line profiles. The obtained
  parameters and their formal errors are $T_\mathrm{eff} =
  10090\pm25$\,K and $\log g = 6.44\pm0.05$\,cgs. The
  resulting best-fit model is shown in Fig.~\ref{fig:5}. With this
  temperature, the spectral type of the white dwarf is DA5
  \citep{wgl+93}.

  The model fits the observed Balmer lines extremely well up to
  H10. H11 and H12 are weaker and not as deep in the model as is
  observed. This may be an indication that the surface gravity is
  slightly lower than the formal fit. It may also indicate that the
  Hummer-Mihalas occupation probability theory
  \citep{hm88,mdh88,mhmd90} used in the models overestimates the
  quenching of the highest Balmer lines. The theoretical line profiles
  of H11 and H12 depend strongly on the Hummer-Mihalas formalism and
  as these lines are absent in the typical $\log g=8$ white dwarf, the
  theory is difficult to calibrate. For a recent study on the
  influence of the Hummer-Mihalas formalism on the line profiles, see
  \citet{knv+05}.

  The fit uncertainties include only the statistical errors, while
  systematic effects may be important \citep{rei96,vska97}. We
  experimented with small changes in the continuum and fit regions,
  and other fitting parameters. The largest effect is from changes in
  the resolution: assuming that the resolution were 6.5\AA\ instead of
  the 4.5\AA\ inferred from the lines in the spectrum of the reference
  star, we find $T_\mathrm{eff}=10135$\,K and $\log g=6.20$\,cgs. As
  we discuss in Appendix~\ref{ssec:b.2.2}, rapid rotation may mimic a
  change in resolution. Since this is not secure, however, we decided
  to adopt conservative errors below, and use $T_\mathrm{eff} =
  10090\pm150$\,K and $\log g = 6.44\pm0.20$\,cgs

  \subsection{White dwarf mass and radius}\label{ssec:4.3}
\begin{figure}
    \resizebox{\hsize}{!}{\includegraphics{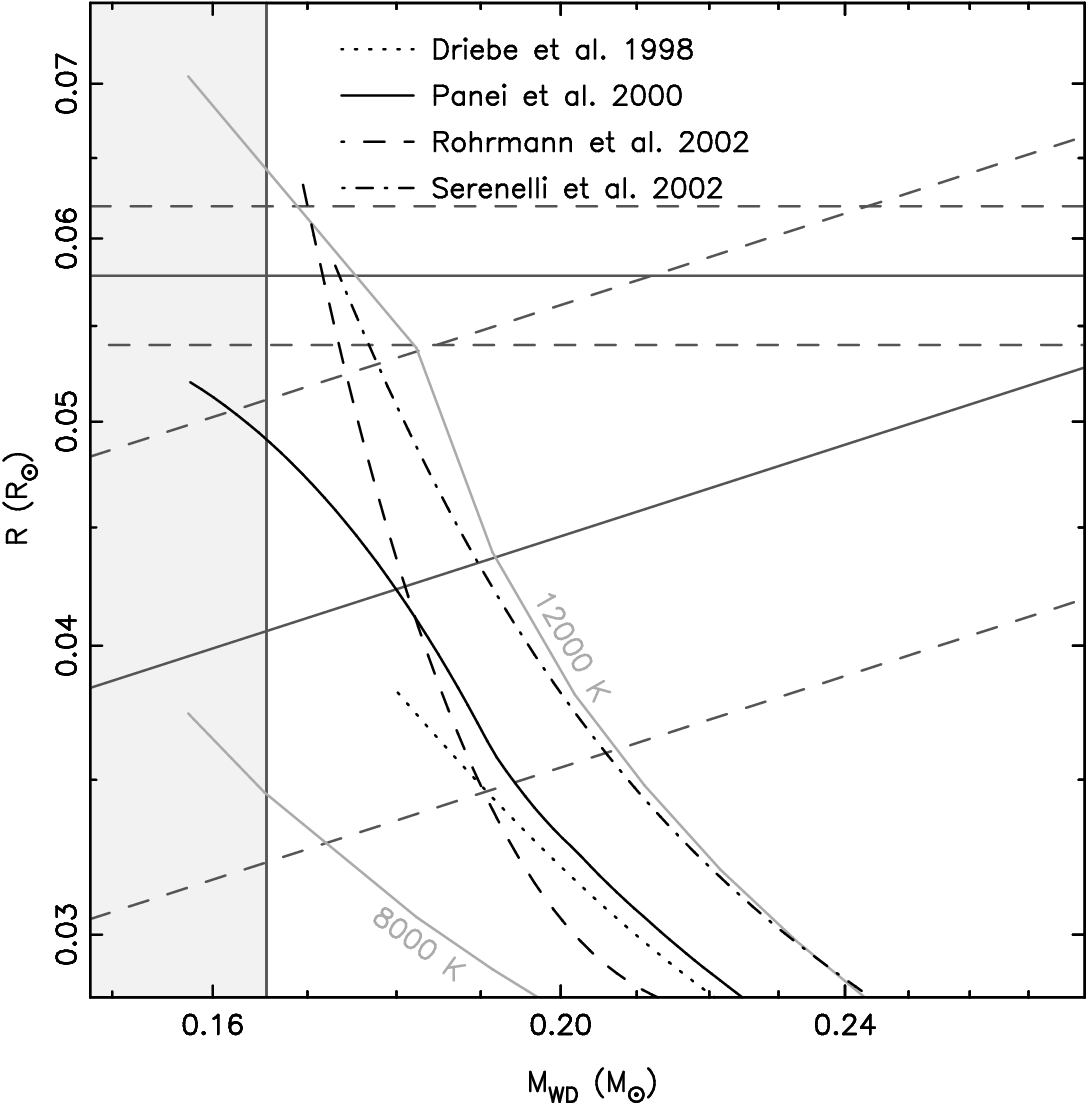}}
    \caption{White-dwarf mass-radius relations for a temperature of
    $T_\mathrm{eff}=10090\pm150$\,K. Shown are the relations from
    \citet{rsab02} (\emph{dashed line}), \citet{sarb02} (\emph{dashed
    dotted}), \citet{dsbh98} (\emph{dotted}) and \citet{pab00}
    (\emph{solid lines}). For the latter model, the 8\,000\,K and
    12\,000\,K models used to obtain the mass-radius relation at the
    observed temperature are also shown. The diagonal solid and dashed
    lines depict the observed range ($1\sigma$) in surface gravity
    ($\log g=6.44\pm0.20$\,cgs, with $g=GM/R^2$). The horizontal solid
    and dashed lines indicate the white-dwarf radius determined using
    the distance of NGC\,6752. The filled light grey area depicts the
    region excluded by the $2\sigma$ lower limit on the white-dwarf
    mass ($M_\mathrm{WD}>0.166\,M_\odot$) that was derived from the
    pulsar mass-function and the observed mass ratio.}
    \label{fig:6}
  \end{figure}

  The mass and radius of the white dwarf can be determined from the
  surface gravity using a mass-radius relation appropriate for a
  helium-core white dwarf at the observed temperature. Such
  mass-radius relations have been modelled by \citet{pab00} and we
  interpolate between their 8000\,K and 12000\,K tracks to obtain the
  relation at the observed temperature, as shown in Fig.~\ref{fig:6},
  giving $M_\mathrm{WD}=0.180\pm0.018\,M_\odot$ and
  $R_\mathrm{WD}=0.0423\pm0.0075\,R_\odot$. Similar white-dwarf masses
  and radii are found from the evolutionary cooling tracks of
  \citet{dsbh98} and \citet{rsab02}. Here we obtain, for each model
  with a given mass, the radius and hence the surface gravity at the
  observed white-dwarf temperature and interpolate between the models
  to get the mass and radius at the observed $\log g$. The models by
  \citet{rsab02} give $M_\mathrm{WD}=0.181\pm0.012\,M_\odot$ with
  $R_\mathrm{WD}=0.0424\pm0.0088\,R_\odot$. The lowest mass model by
  \citet{dsbh98} has 0.179\,M$_\odot$, so we extrapolate their models.
  This yields $M_\mathrm{WD}=0.172\pm0.018\,M_\odot$ and
  $R_\mathrm{WD}=0.0414\pm0.0074\,R_\odot$, though the uncertainties
  may be underestimated because of the extrapolation. The
  uncertainties of the \citet{rsab02} models are considerably smaller,
  as its mass-radius relation is steeper than those of the
  \citet{pab00} and \citet{dsbh98}. Finally, the mass-radius relations
  from the $Z=0.0010$ and $Z=0.0002$ models by \citet{sarb02} are very
  similar and both predict a somewhat higher white-dwarf mass, of
  $M_\mathrm{WD}=0.190\pm0.015\,M_\odot$, and a radius of
  $R_\mathrm{WD}=0.0434\pm0.0084\,R_\odot$.

  We should note that the models by \citet{dsbh98} and \citet{rsab02}
  are computed for white-dwarf progenitors with solar metallicity and
  are appropriate for field systems. As the metallicity of \gc\ is
  considerably smaller ([Fe/H]$=-1.43\pm0.04$, \citealt{gbc+03}), the
  \citet{sarb02} models for white-dwarf progenitors with sub-solar
  metallicities (with $Z=0.0010$ in particular) are more appropriate
  in the case that \psr\ is associated with \gc. The differences of
  about 0.01\,M$_\odot$ and 0.001\,R$_\odot$ in the predictions from
  different models, however, are similar in magnitude to the
  difference induced by different metallicities. For the remainder of
  the paper, we will use values that encompass all predictions from
  the effective temperature and surface gravity:
  $M_\mathrm{WD}=0.18\pm0.02\,M_\odot$ and
  $R_\mathrm{WD}=0.043\pm0.009\,R_\odot$.

  \subsection{Distance inferred from the white dwarf}\label{ssec:4.4}
  The distance to the white dwarf can now be estimated using the
  observed and modelled flux and the radius of the white dwarf.  We do
  this by writing the flux normalization $f=(R/d)^2 \pi F$ between the
  observed flux $f$, the model flux $\pi F$, and the radius $R$ over
  the distance $d$ in terms of magnitudes; $M_\lambda=43.234-5\log
  R/R_\odot-2.5\log \pi F_\lambda+c_\lambda$, where $M_\lambda$ is the
  absolute magnitude in a given filter, $\pi F_\lambda$ the flux from
  the model in the same filter and $c_\lambda$ the zero-point of the
  filter. By convolving the flux-calibrated best-fit model of the
  observed spectrum with the $B$ and $V$-band filter curves from
  \citet{bes90}, we obtain $\pi
  F_B=9.45\times10^7$\,erg\,cm$^{-2}$\,s$^{-1}$\,\AA$^{-1}$ and $\pi
  F_V=6.26\times10^7$\,erg\,cm$^{-2}$\,s$^{-1}$\,\AA$^{-1}$. The
  uncertainties on these fluxes due to the uncertainties in
  $T_\mathrm{eff}$ and $\log g$ are about 5\%. With the zero-points
  from \citet{bcp98}, $c_B=-20.948$ and $c_V=-21.100$, we obtain
  $M_B=9.63\pm0.46$, $M_V=9.48\pm0.46$. The uncertainties are
  dominated by those on the white-dwarf radius.

  Combined with the observed $B$ and $V$-band magnitudes from
  Table~\ref{tab:2}, we obtain distance moduli of
  $(m-M)_B=12.59\pm0.46$ and $(m-M)_V=12.66\pm0.46$. Assuming a
  negligible reddening, the averaged distance modulus yields a
  distance of $d=3.4\pm0.7$\,kpc. Assuming a reddening of
  $E_{B-V}=0.05$, the distance becomes $d=3.1\pm0.7$\,kpc.

  \subsection{Mass and radius from distance, flux and temperature}\label{ssec:4.5}
  Now we determine the properties of the white dwarf based on the
  assumption that the distance of the binary is that of the globular
  cluster. From the values from Table~\ref{tab:2}, the distance
  modulus $(m-M)_V=13.24\pm0.08$ and reddening $E_{B-V}=0.046\pm0.005$
  \citep{gbc+03,gbc+05}, we obtain $M_B=8.93\pm0.09$ and
  $M_V=8.89\pm0.08$. The $B$ and $V$-band fluxes and zero-points and
  the relation from \S~\ref{ssec:4.4} yield white-dwarf radii of $\log
  R/R_\odot=-1.226\pm0.021$ and $\log R/R_\odot=-1.249\pm0.018$ for
  the $B$ and $V$-band values, respectively. As a conservative
  estimate, we will use $\log R/R_\odot=-1.238\pm0.030$, corresponding
  to $R=0.058\pm0.004\,R_\odot$, which encompasses both values.

  The mass of the white dwarf can be determined from the radius using
  the mass-radius relations. The radius is shown with the horizontal
  lines in Fig.~\ref{fig:6}, together with the mass-radius
  relations. The models by \citet{rsab02} predict
  $M_\mathrm{WD}=0.172\pm0.001\,M_\odot$, while the models of
  \citet{sarb02} give $M_\mathrm{WD}=0.175\pm0.002\,M_\odot$.  The
  mass-radius relation by \citet{pab00} and the evolutionary models by
  \citet{dsbh98} do not reach these radii and provide no mass
  estimate. The surface gravity that corresponds with these masses and
  radii is about $\log g=6.20$\,cgs. Both the white-dwarf
  radius and the surface gravity are slightly outside the $1\sigma$
  range of these values inferred from the spectrum (see
  Fig.~\ref{fig:6}). Because of the steepness of the mass-radius
  relations, the masses are in agreement. We do note that if the
  pulsar binary is at the distance of \gc, the white-dwarf mass is
  slightly below our best estimate for the minimum mass, though they
  are consistent with the $2\sigma$ limit.

  If the pulsar binary is associated with \gc, the progenitor of the
  white dwarf must have evolved from a cluster star. \citet{sarb02}
  computed white-dwarf cooling tracks for objects with sub-solar
  metallicities. Their $Z=0.001$ models have a metallicity that is
  similar to that of \gc\ ([Fe/H]$=-1.43\pm0.04$; \citealt{gbc+03}),
  and for this reason, we use the corresponding mass estimate as the
  best value for the white-dwarf mass. To take into account the
  uncertainty in the white-dwarf mass-radius relation, we add
  0.010\,M$_\odot$ in quadrature to the uncertainty in the mass. In
  summary, under the assumption that the binary system is a member of
  \gc, we infer a radius of $R_\mathrm{WD}=0.058\pm0.004\,R_\odot$ and
  a mass of $M_\mathrm{WD}=0.175\pm0.010\,M_\odot$.
  
  \begin{figure}
    \resizebox{\hsize}{!}{\includegraphics{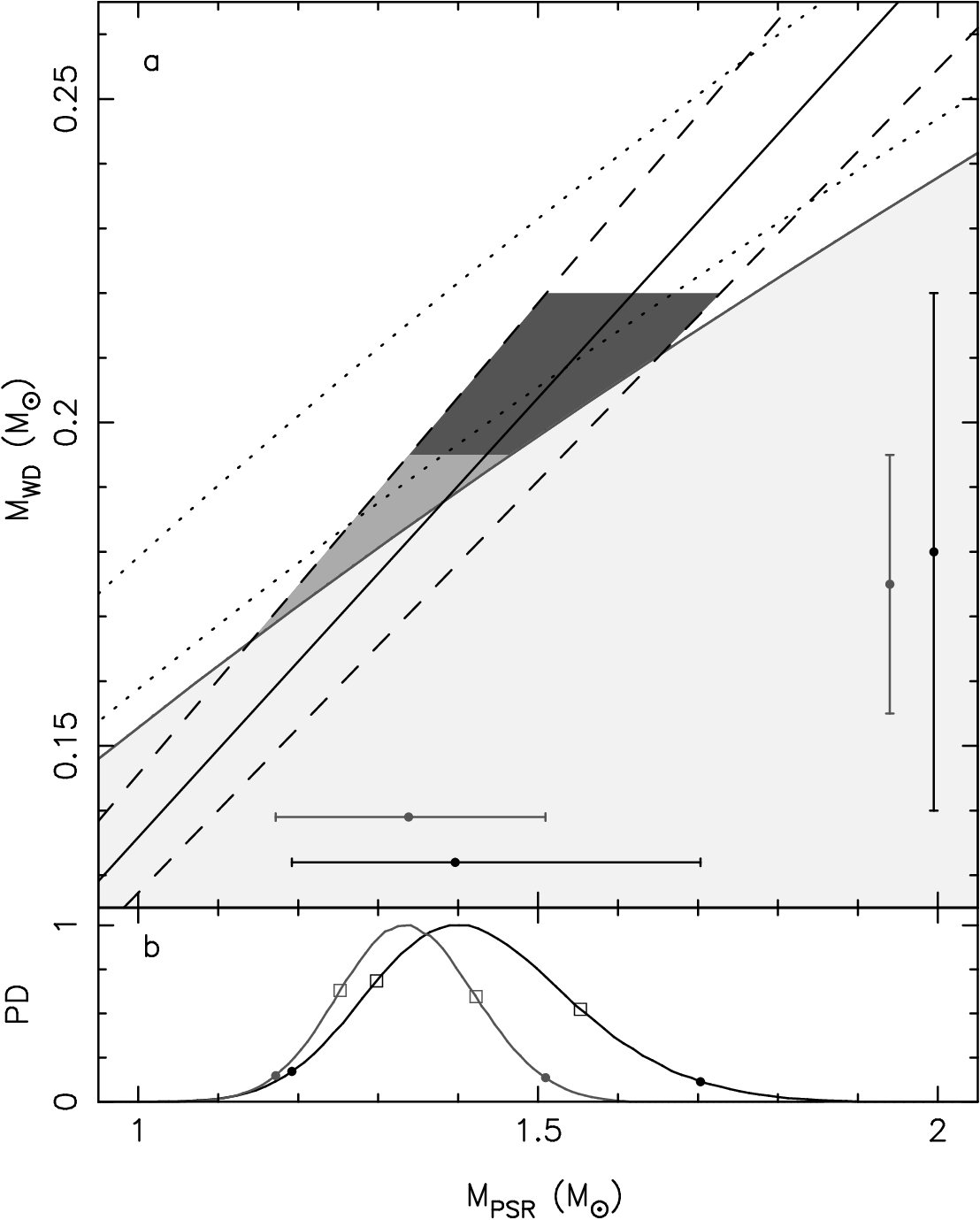}}
    \caption{The various constraints on the mass of
      \psr. Panel~\textbf{a} shows the constraint set by the mass
      ratio as the diagonal solid and dashed black lines. The large
      filled light grey area is excluded by the mass-function of the
      pulsar, as points in this area would require $\sin i>1$. The
      line that borders this area is for the limit $i=90\degr$ in the
      pulsar mass-function. The two dotted lines depict inclinations
      of $i=75\degr$ (lower line) and $i=60\degr$ (upper line). Two
      pairs of vertical error bars on the right-hand side of the panel
      represent the two white-dwarf mass estimates and their
      uncertainties, where the larger error bar is for the mass
      determined from the surface gravity and effective temperature,
      whereas the smaller error bar is from the assumption that the
      binary is a member of \gc. Allowed values for the pulsar mass
      exist in the light grey region for a white dwarf associated with
      \gc, and the light and dark grey area for a white dwarf not
      associated with the globular cluster. The resulting pulsar
      masses, based on the Monte Carlo simulation, are indicated with
      the horizontal error bars.  The uncertainties on the white-dwarf
      mass, pulsar mass, mass ratio and the mass-function are all 95\%
      confidence ($2\sigma$). The probability densities as a function
      of pulsar mass are shown in panel~\textbf{b}. The 68\% and 95\%
      confidence uncertainties based on these distributions are
      denoted by the open squares and the filled circles,
      respectively.}
    \label{fig:7}
  \end{figure}
  
  \subsection{Pulsar mass}
  As for the white dwarf, we can determine a minimum mass for the
  pulsar from the mass ratio and the constraint that the inclination
  $i$ is less than or equal to 90\degr. The white-dwarf mass-function
  is $f(M_\mathrm{WD})=M_\mathrm{PSR}^3 \sin^3
  i/(M_\mathrm{WD}+M_\mathrm{PSR})^2=K_\mathrm{WD}^3
  P_\mathrm{b}/(2\pi G)=1.072\pm0.108\,M_\odot$, which can be written
  to give $M_\mathrm{PSR}\sin^3 i=(1+1/q)^2 f(M_\mathrm{WD})$. With
  $\sin i\le1$, this yields a $1\sigma$ lower limit to the pulsar
  mass: $M_\mathrm{PSR}>1.24\,M_\odot$. The $2\sigma$ limit is
  $M_\mathrm{PSR}>1.10\,M_\odot$. These limits are model independent,
  as they only depend on three observables; the projected semi-major
  axis of the pulsar orbit $a_\mathrm{PSR} \sin i$ and the orbital
  $P_\mathrm{b}$, as determined from pulsar timing, and the
  white-dwarf velocity amplitude $K_\mathrm{WD}$, that we obtained
  from our spectroscopic observations.

  In Fig.~\ref{fig:7}, we show the constraints set by the mass ratio
  and the two white-dwarf mass determinations. It is clear that a
  large part of the range allowed by these constraints is excluded by
  the pulsar mass function. As a result, the most probable value for
  the pulsar mass and the uncertainties on it are not normally
  distributed. Instead, we determined these values via a Monte Carlo
  error propagation method. For a million trial evaluations, values
  for $P_\mathrm{b}$, $a_\mathrm{PSR}\sin i$, $K_\mathrm{WD}$ and
  $M_\mathrm{WD}$ were randomly drawn from Gaussian distributions with
  the appropriate means and widths to obtain the corresponding pulsar
  mass. Solutions that had $\sin i>1$ were excluded. From the
  resulting distribution of solutions the most probable value for and
  the uncertainties on the pulsar mass were determined.

  For a white-dwarf mass of $M_\mathrm{WD}=0.18\pm0.02\,M_\odot$, the
  mass of the pulsar is constrained to
  $M_\mathrm{PSR}=1.40^{+0.16}_{-0.10}\,M_\odot$ at 68\%
  confidence. For the case that the white dwarf is associated with the
  globular cluster, the allowed range in pulsar mass is smaller,
  $1.34\pm0.08\,M_\odot$ at 68\% confidence. The uncertainties
  corresponding to 95\% confidence are in both cases twice as large.

  \section{Discussion and conclusions}\label{sec:5.0}
  We have unambiguously identified the companion to \psr\ as a
  Helium-core white dwarf and determined its mass. Together with the
  measurement of the mass-ratio of the binary, we obtain constraints
  on the pulsar mass. However, before we discuss our results, we
  compare our results to those presented by \citet{cfpd06}.

  Cocozza et al.\ found that the light-curve of the white dwarf
  companion to \psr\ showed two phases of brightening by about
  0.3\,mag, located approximately at the quadratures of the orbit
  (phases $\phi=0.0$ and $\phi=0.5$). This result is at odds with our
  light-curve, which excludes variations larger than
  0.02\,mag. Fig.\,\ref{fig:3} shows a reproduction of the light-curve
  determined by \citet{cfpd06}. This figure also shows the variation
  in the seeing under which these images were obtained; one sees that
  these closely follow the variations in the white-dwarf magnitude
  found by \citet{cfpd06}. This suggests that seeing affects their
  photometry and that the variations they measure are due to
  variations in the seeing and not due to variations in the
  white-dwarf brightness. We found that we could reproduce their
  light-curve by defining the PSF over an area smaller than about
  $3\farcs1$ in radius. This distance corresponds to the distance
  between star R and the white dwarf, and if the PSF radius is chosen
  smaller than this value, flux in the wings of star R is not removed
  and added to the flux of the white dwarf. Hence, the effect
  increases for increasing seeing. We used a PSF radius that extends
  up to $4\arcsec$ from the center of each star and is still
  $2.5\times$ larger than the width of the PSF in the images of the
  worst seeing. As such, our photometry is not affected by this error
  and excludes the 0.3\,mag variations seen by \citet{cfpd06}.

  The radial-velocity curve of the white dwarf companion to \psr\ is
  also presented in \citet{cfpd06}. Their radial-velocity amplitude
  $K_\mathrm{WD}$ and systemic velocity $\gamma$ are consistent with
  the values we found using the uncorrected velocities. We note that
  the uncertainty on our value for $K_\mathrm{WD}$ is about a factor
  two smaller. We believe this is caused by the fact that we used nine
  Balmer lines (H$\beta$ up to H12) whereas \citet{cfpd06} only used
  four (H$\beta$ up to H$\epsilon$). Especially since the higher
  Balmer lines are narrower they will provide more accurate
  velocities.  \citet{cfpd06} use their measurement of the systemic
  velocity of the pulsar binary ($\gamma=-28.1\pm4.9$\,\kms) as an
  arguement supporting the association of \psr\ with \gc. However, as
  \citet{cfpd06} did not correct for the systematic shifts in the
  wavelength scale that we identified and corrected for (see
  Appendix~\ref{sec:a}), their conclusion regarding the association
  between the binary and the cluster is meaningless.

  We now turn to the conclusions that can be drawn from our results.
  In Appendix\,\ref{sec:b} we have used the available constraints set by
  our observations to test whether \psr\ is associated with the
  globular cluster \gc. Unfortunately, these tests are not conclusive
  and hence, we discuss both pulsar mass determinations below.

  First, it is interesting to compare the mass of the white dwarf with
  the mass predicted by the theoretical relation between the
  white-dwarf mass and the orbital period \citep{jrl87}. For short
  orbital periods, this relation is least secure, since mass transfer
  starts before the companion has a well-developed core
  \citep{esa98}. Nevertheless, from earlier systems it seemed that the
  predictions by \citet{ts99}, which are strictly valid only for
  $P_\mathrm{b}>2$\,d, work well for binaries with orbital periods as
  short as 8\,h (see Fig.~2 in \citealt{kbjj05}). At the orbital
  period of \psr, their relation predicts a white-dwarf mass between
  0.18 and 0.20\,$M_\odot$. This is again in very good agreement with
  our white-dwarf mass measurement (independent of whether the pulsar
  binary is associated with the globular cluster).

  Binary evolution furthermore predicts that the progenitors of white
  dwarfs in low-mass binary millisecond pulsars have lost
  $\ga\!0.6\,M_\odot$ of matter in order to form a $\sim\!0.2\,M_\odot$
  helium-core white dwarf. It is believed that at least a part of this
  matter is accreted onto the neutron star in order to spin it up to
  millisecond periods. As such, the neutron stars in low-mass binary
  pulsar systems are expected to be heavier than the canonical neutron
  star of $1.35\pm0.04\,M_\odot$ \citep{tc99}.

  For the case that \psr\ is a field system, the mass of the pulsar
  ($M_\mathrm{PSR}=1.40^{+0.16}_{-0.10}\,M_\odot$) is indeed heavier
  than the canonical value, though not by much. However, similarly
  small differences have been found for several of the other low-mass
  binary millisecond pulsars for which masses have been determined;
  PSR~J1713+0747, $1.3\pm0.2\,M_\odot$ \citep{sns+05},
  PSR~J1909$-$3744, $1.438\pm0.024\,M_\odot$ \citep{jhb+05},
  PSR~J0437$-$4715 with $1.58\pm0.18\,M_\odot$ \citep{sbb+01},
  PSR~B1855+09, $1.6\pm0.2\,M_\odot$ \citep{nss05} PSR~J1012+5307,
  $1.6\pm0.2\,M_\odot$ \citep{kbk96,kbjj05,cgk98}.  The only system
  for which the pulsar is significantly heavier than the
  $1.35\,M_\odot$ is PSR~J0751+1807, with $2.1\pm0.2\,M_\odot$
  \citep{nss+05}.

  For the case that \psr\ is associated with \gc, the pulsar mass
  ($1.34\pm0.08\,M_\odot$) is consistent with the
  $1.35\pm0.04\,M_\odot$ found by \citet{tc99} and is one of the least
  heavy neutron stars in low-mass binary millisecond pulsars. In this
  case, it is interesting to compare \psr\ with PSR~J0737$-$3039B, the
  2.8\,s non-recycled pulsar in the double pulsar system. If the mass
  of this pulsar ($1.250\pm0.005\,M_\odot$; \citealt{lbk+04}) is
  indicative of the mass of a neutron star after is has been formed,
  it would only take less then $0.1\,M_\odot$ to recycle a normal
  neutron star to a millisecond pulsar spinning with a period of
  3.26\,ms.

  Finally, our observations also constrain the inclination of the
  system. For the case that \psr\ is not associated with \gc, we have
  a $2\sigma$ limit of $\sin i>0.923$ or $i>67\fdg4$. In the other
  case, $\sin i>0.968$ and $i>75\fdg5$. Because of these high
  inclinations, the effects of Shapiro delay should be significant in
  the timing of the pulsar. Combined with our measurement of the
  white-dwarf mass, these limits on the inclination imply a Shapiro
  delay $\Delta_S>5.7\,\mu$s. Unfortunately, for nearly circular
  orbits, the Shapiro delay is covariant with the projected semi-major
  axis and the eccentricity, and a large part of the effect is
  absorbed in these two parameters. As a result, the effect that
  remains has a size of ${\Delta_S}'>1.2\,\mu$s. Interestingly, if
  Shapiro delay is present, but neglected in the pulsar timing fit, it
  introduces an apparent eccentricity of $e>1.3\times10^{-6}$ and
  places the longitude of periastron at $\omega=90\degr$. Though the
  small signal due to Shapiro delay may be difficult to detect, radio
  timing observations of \psr\ may be used to obtain an upper limit on
  the inclination and the companion mass. For example, if $i=85\degr$,
  the timing signal due to Shapiro delay will be much larger,
  $\Delta_S=11.2$\,$\mu$s and ${\Delta_S}'=5.4$\,$\mu$s.

  \begin{acknowledgements}
    This research is based on observations made with ESO Telescopes at
    the Paranal Observatory under programme IDs 71.D-0013, 71.D-0232,
    073.D-0039 and 073.D-0067. We thank Daniel Harbeck for providing
    us with proprietary VLT/FORS1 data.  MIDAS is developed and
    maintained by the European Southern Observatory. We acknowledge
    support from NWO (C. G. B.) and NSERC (M. H. v. K.).
  \end{acknowledgements}


  \appendix

  \section{Corrections to the wavelength calibration}\label{sec:a}
\begin{figure}
    \resizebox{\hsize}{!}{\includegraphics{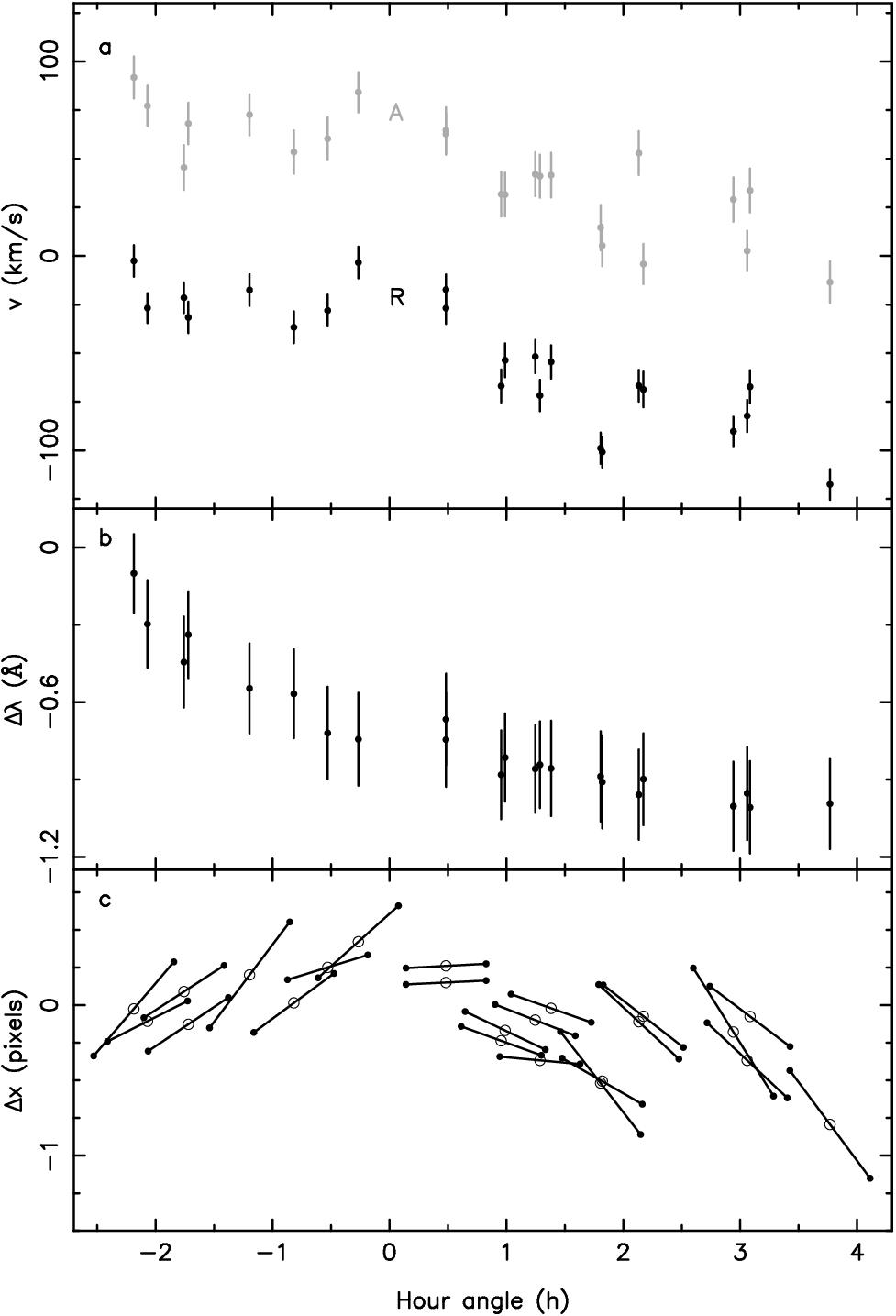}}
    \caption{The systematic effects that were present in the radial
    velocity study. \textbf{a} The uncorrected radial velocities of
    stars R and A as a function of hour angle. For both stars, the
    velocities decrease with increasing hour angle, while the scatter
    around the main trend is comparable for both stars. \textbf{b} The
    offset in the measured position of the \ion{O}{i} $\lambda5577$
    sky line compared to the laboratory value. \textbf{c} The
    centering of star R on the slit, as measured from the 'before' and
    'after' through-the-slit images. Each line connecting the two
    black dots correspond to a single spectrum, where the left dot is
    the position determined in the 'before' image, and the right-hand
    dot the position in the 'after' image. The open circle is the
    average of the two positions in hour angle and pixel shift.}
    \label{fig:a1}
  \end{figure}

  First, according to the wavelength calibration of the different
  spectra, the wavelength of the \ion{O}{i} $\lambda5577$ night sky
  emission line was offset from the laboratory value
  ($\lambda=5577.34$\,\AA). The offsets varied over a range of 0.9\AA\
  or about 50\,\kms\ and appeared to decrease with increasing hour
  angle, as shown in Fig.~\ref{fig:a1}b. We believe this effect is the
  result of instrument flexure due to differences in the pointing of
  the telescope, as the calibration frames are obtained with the
  telescope pointing towards the zenit, while the \psr\ has $\sec
  z>1.24$. We corrected for this effect by applying the measured
  offsets as a wavelength shift in the zero point of the wavelength
  calibrations of each individual spectrum.

  However, as the \ion{O}{i} $\lambda5577$ sky line lies redwards of
  the last arc line used for the wavelength calibration (\ion{Hg}{}
  $\lambda5460$) we cannot verify that the polynomial used for the
  wavelength calibration is still valid at the wavelength of the
  skyline and that all systematic wavelength shifts are corrected
  for. We found, however, that the higher order terms of the
  polynomial are very similar between the calibration spectra
  extracted for the different stars; this is because the curvature of
  the calibration lines and the sky line at the positions of the stars
  is very small. As a result, we do not expect random effects to
  dominate the wavelength shift computed above, though the procedure
  may introduce a systematic wavelength shift affecting all spectra.

  The second effect involved the centering of the stars on the
  $1\farcs31$ slit. In general, when a star is not centered on the
  slit the offset of the star from the center of the slit will appear
  as a wavelength shift in the spectrum. This can clearly be seen by
  the velocities of star R and A in Fig.\,\ref{fig:a1}a, where the
  velocity of star A differs from that of star R by about 100\,\kms;
  this is primarily due to the fact that compared to star R, star A is
  positioned closer to the right-hand (redwards) edge of the slit, see
  Fig.\,\ref{fig:1}. Because of the magnitude of this effect, we
  determined the position of the reference star R with respect to the
  center of the slit from the through-the-slit images taken before and
  after the spectra. The positions on the `before' images varied over
  a range of 0.68\,pixels, whereas those on the `after' images varied
  over 1.81\,pixels. Interestingly, the `after' positions with
  negative hour angles displayed positive shifts (redwards in
  wavelength) whereas those with positive hour angles displayed
  negative shifts. Fig.\,\ref{fig:a1}c shows these trends. We interpret
  the variations in centering of the reference star R on the `before'
  images as simple scatter inherent to the centering of a star on the
  slit. The centering variations on the `after' images clearly has a
  different cause, given its dependence on the hour angle. It may be
  that this is also related to flexure or differential atmospheric
  refraction. The averaged affect introduces the scatter in the
  velocities of the stars on the slit (Fig.\,\ref{fig:a1}a) and affects
  all stars in the same way.

  To correct for this effect, we computed the position $\bar{x_i}$ of
  star $i$ (in pixels) with respect to the center of the slit and
  applied it as a shift in wavelength in the zeropoint of the
  wavelength calibrations of each individual spectrum. The position is
  computed from $\bar{x_i} = \int_{-w/2}^{w/2} x\
  \psi_\mathrm{PSF}(x-x_\mathrm{R}-\Delta x_i)\ \mathrm{d} x$, where
  $\psi_\mathrm{PSF}$ is the normalized point-spread function, as
  determined from fitting a Moffat function to the spatial stellar
  profile. Furthermore, $x_\mathrm{R}$ is the average of the `before'
  and `after' shifts in the position of the reference star R with
  respect to the center of the slit and $w$ is the slit width (in
  pixels). Finally, $\Delta x_i$ is the offset between the position of
  the reference star R and star $i$, measured in pixels in the
  dispersion direction. These offsets were computed from the five
  combined 360\,s $B$-band exposures used for the photometry.

  The resulting radial velocities of the pulsar companion and the
  other stars on the slit, corrected to the solar-system barycenter,
  are given in Table~\ref{tab:1}. The velocity of the reference star R
  now only varies over a range of 30\,\kms, with an rms scatter of
  11\,\kms, comparable to the errors on the velocities tabulated in
  Table~\ref{tab:1}. For stars A, B, C, and D the rms scatter is
  comparable or somewhat larger, with 10, 50, 32 and 13\,\kms,
  respectively. The large scatter in the radial velocities of star B
  is likely caused by secondary light from the two nearby stars (see
  Fig.~\ref{fig:1}), where variations in the seeing lead to shifts in
  the center of light in the dispersion direction, which in turn leads
  to velocity shifts. Therefore, we have not used star B in the
  further analysis.

  The stars on the slit have systemic radial velocities of
  $\gamma_\mathrm{R}=8\pm2$\,\kms, $\gamma_\mathrm{A}=7\pm2$\,\kms,
  $\gamma_\mathrm{C}=30\pm2$\,\kms and
  $\gamma_\mathrm{D}=6\pm2$\,\kms. It is unexpected that all these
  stars have systemic velocities that are different from the radial
  velocity of the globular cluster \gc\ ($-32.0\pm1.6$\,\kms;
  \citealt{dmm97}), especially as all stars, except star~C, have
  colours that place them on the narrow cluster main sequence in
  Fig.~\ref{fig:2}. Since star R, A and D have very similar systemic
  velocities, we conclude that these stars are cluster members, but
  that their velocities are off by $-39\pm3$\,\kms. We suspect that
  this systematic shift may have been introduced by the correction
  that we applied using the night-sky emission line. As this sky line
  was located redwards of the reddest wavelength calibration line, the
  polynomial could have introduced this systematic shift. However, as
  mentioned above, we do not expect that this influences relative
  velocities between different observations and different stars, since
  the shape of the polynomial did not vary between the different
  observations.

\begin{figure}
    \resizebox{\hsize}{!}{\includegraphics{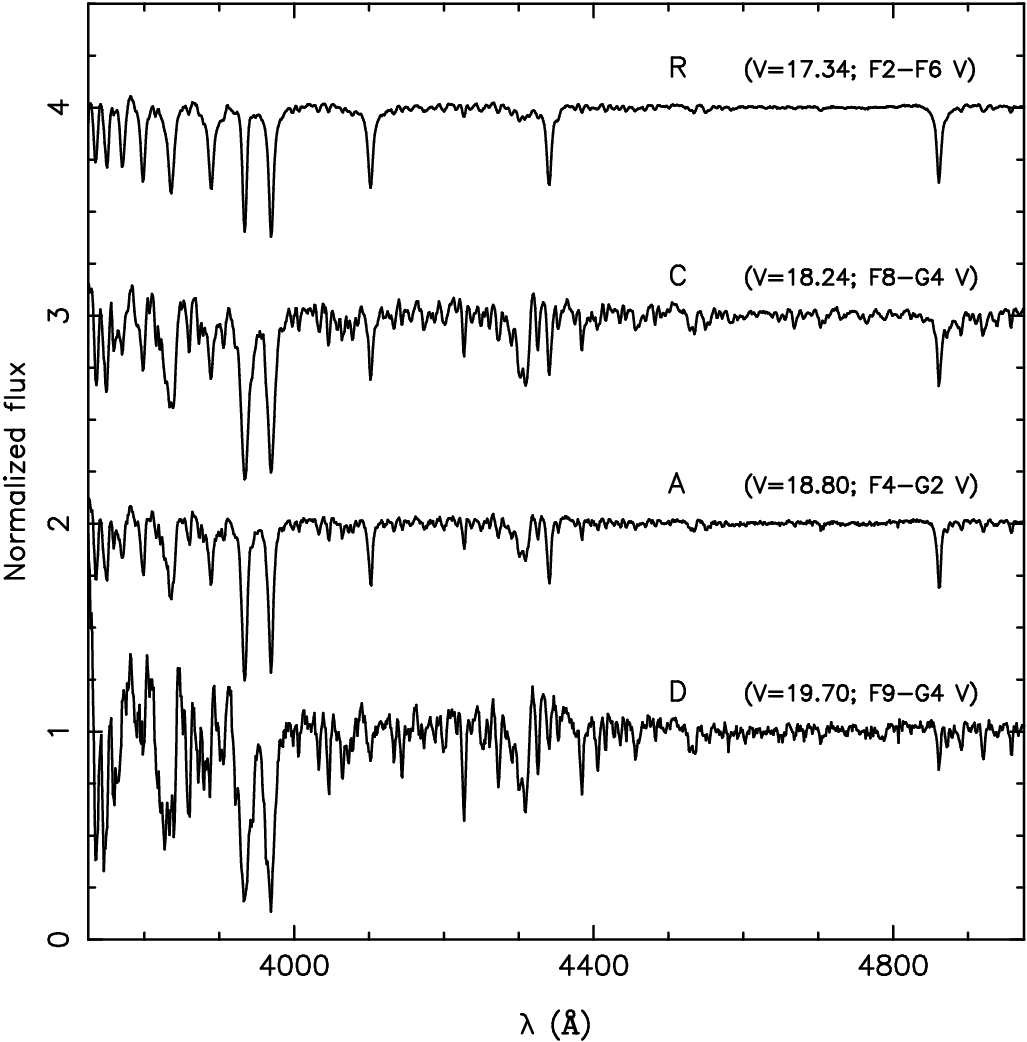}}
    \caption{Normalized spectra of the stars on the slit. The spectra
      are ordered from the star with the lowest $V$-band magnitude, star
      D, to the brightest, star R.  Each consecutive spectrum is shifted
      by one unit.}
    \label{fig:a2}
  \end{figure}

  The one remaining issue is that of the systemic radial velocity of
  star C, which is different from that of the other stars. Star C may
  not be a cluster member, as it does not coincide with the cluster
  main sequence. Furthermore, its spectral features, as displayed in
  Fig.~\ref{fig:a2}, are distinct from those of the other stars. In
  this figure, the normalized spectra of the stars are sorted in order
  of increasing $V$-band magnitude, so that when the stars are cluster
  members their spectra should be ordered on spectral type and their
  spectral features should change accordingly. Approximate spectral
  types for these stars were determined by comparing the spectra to
  those in the atlas of \citet{bbp+03}. We see that both the spectral
  features and the spectral type of star C show more resemblance with
  that of star D than they do to star R or A. From this we conclude
  that star C is not a cluster member and therefore its radial
  velocity may differ from that of the cluster.

  \section{Is the binary associated with \gc?}\label{sec:b}
  We use our observations and the properties of the white
  dwarf and the pulsar that we derived from them, to test if these
  support the assumption that \psr\ is a member of the globular
  cluster \gc.

  \subsection{Velocities}\label{ssec:b.1}
  If the pulsar binary is a member of the globular cluster, the
  systemic radial velocity of the binary should be compatible with the
  radial velocity of the cluster, $-32.0\pm1.6$\,\kms\ \citep{dmm97}.

  The absolute systemic velocity of the pulsar binary is
  $\gamma=-18\pm6$\,\kms. This differs from the radial velocity of the
  cluster by $14\pm6$\,\kms\ and is consistent with the systemic
  velocity of the binary $\gamma_{\Delta v}=13\pm6$\,\kms\ which we
  determined from the fit of the white dwarf velocities relative to
  those of star R.

  The velocity difference needs to be corrected for the gravitational
  redshift of the white dwarf and the reference star R. Using the mass
  and radius of the white dwarf determined in \S~\ref{ssec:4.3}, and
  the mass and radius of star~R determined from the its absolute
  magnitude and the models by \citet{gbbc00} plotted in
  Fig.~\ref{fig:2}; we obtain 2.7\,\kms\ and 0.5\,\kms\ for the white
  dwarf and star R, respectively. This reduces the velocity difference
  to $11\pm6$\,\kms, amounting to about a $2\sigma$ difference between
  the velocity of the binary and the velocity of the cluster. A part
  of this velocity difference may be due to the dispersion in the
  velocity between the pulsar binary and the reference star R. From
  the scatter in the velocities of stars R, A and D, we estimate that
  the velocity dispersion is of the order of $\sim\!2$\,\kms.

  Also important is the local escape velocity at the pulsar position
  relative to the cluster center. To estimate the escape velocity
  $v_\mathrm{esc}=2GM/r$ at the projected distance $r_\perp$ of \psr,
  we compute the mass $M(r_\perp)$ inside a sphere of radius $r_\perp$
  using relation A3 from the simplified cluster model presented in the
  Appendix of \citet{fhn+05}. Here, we used a core-radius of
  $r_\mathrm{c}=6\farcs7$ \citep{lcg95}, a distance of $d=4.14$\,kpc
  \citep{gbc+03}, a central velocity dispersion of
  $\sigma_z(0)=4.5$\,\kms\ \citep{dmm97} and $r_\perp=6\farcm37$ to
  obtain $M(r_\perp)=27\times10^3\,M_\odot$ and
  $v_\mathrm{esc}=5.7$\,\kms. This velocity should be regarded as an
  upper limit since only the projected distance to the cluster center
  is known and not the actual distance $r^2=r_\perp^2+r_{||}^2$, with
  $r_{||}$ the distance along the line-of-sight towards \gc.
  
  We conclude that, taking into account the velocity range expected
  due to the velocity dispersion and the escape velocity, the systemic
  velocity of \psr\ is only marginally consistent (at the $2\sigma$
  level) with the radial velocity of \gc. 

  \subsection{The radius of the white dwarf}\label{sec:b.2}
  From \S~\ref{sec:4.0}, we found that the white-dwarf radius
  determined from the effective temperature and surface gravity is
  outside the $1\sigma$ uncertainty on the radius derived if the white
  dwarf is assumed to be at the distance of \gc. This suggests that
  the pulsar binary is not a member of the globular cluster. There may
  be additional uncertainties in some of the parameters that were
  used. Here we discuss some of the sources of uncertainty that may
  explain the discrepancy in the white-dwarf properties.

  \subsubsection{Distance, reddening and magnitudes}\label{ssec:b.2.1}
  As a result of the discrepancy in the white-dwarf radius, there is a
  discrepancy in the distance to the white dwarf. The distance modulus
  $(m-M)_V=12.66\pm0.46$ derived from the combination of the
  atmospheric properties of the white dwarf and the mass-radius
  relation (\S~\ref{ssec:4.4}) is only marginally consistent with the
  distance modulus $(m-M)_V=13.24\pm0.08$ \citep{gbc+03} determined
  for \gc.  Though there is a considerable spread in the distance
  modulus measurements, ranging from $(m-M)_V=13.17\pm0.13$
  ($d=4.0\pm0.3$\,kpc) from white-dwarf cooling sequence fitting
  \citep{rbf+96}, to $(m-M)_V=13.32\pm0.04$ ($d=4.31\pm0.08$\,kpc)
  from main-sequence fitting \citep{cgcf00}, no distance determination
  places \gc\ so close. Similarly, the spread in reddening
  measurements, $E_{B-V}=0.035$ to 0.046 \citep{gbc+03,gbc+05}, does
  not help to explain the radius discrepancy.

  It also seems unlikely that our photometry is in error by as much as
  the 0.6\,mag or more needed to match the distance and hence the
  white-dwarf radius. Our \ubv\ magnitudes are consistent with those
  given by \citet{fpsd03}, but have significantly smaller
  uncertainties. The presence of photometric $B$ and $V$ standards on
  the science images removed any uncertainties on the airmass
  dependence. Also, the $B$-band calibrations taken on two different
  nights were consistent with each other, having only a 0.01\,mag
  difference. Finally in \S~\ref{ssec:2.2}, we found that the
  white-dwarf companion is not variable (down to 0.02\,mag) and this
  eliminates the possibility that the \ubv\ photometry was taken at an
  extremum in white-dwarf brightness.

  \subsubsection{Line broadening}\label{ssec:b.2.2}
  As was found in \S~\ref{ssec:4.2}, a lower surface gravity, which
  would imply a lower mass and a larger radius, was found when the
  spectral resolution was decreased to 6.5~\AA. Though this is
  considerably larger than the 4.5~\AA\ determined from the width of
  the lines in the spectrum of star~R, the Balmer lines in the
  spectrum of the white dwarf may be broadened. In order to broaden
  the lines from 4.5~\AA\ to 6.5~\AA\ a velocity smearing of
  $\sim\!300$~\kms\ is required.
  
  One source of broadening is due to the fact that in a single
  observation, the 2470\,s exposure time covers about $\Delta
  \phi=3.5$\% of the 20\,h orbit. This introduces a maximum change
  in velocity (at $\phi=0.25$ and $\phi=0.75$) of about
  $2\pi\Delta\phi K_\mathrm{WD}\simeq50$\,\kms, which is much less
  than required\footnote{It also causes a reduction in inferred
  radial-velocity amplitude by a factor $\sin (\pi \Delta\phi)/\pi
  \Delta\phi=0.998$. This is sufficiently small that we have chosen to
  ignore it.}.

  Another source of broadening could be due to rotation. To estimate
  the rotational velocity of the white dwarf, we follow the reasoning
  used in \citet{kk95} to explain the variations seen in the spectrum
  of the white-dwarf companion to PSR~B0655+64. During the period of
  mass transfer, the progenitor of the white dwarf, a giant, filled
  its Roche lobe and tides ensured the system was synchronized and
  circularised. When mass transfer ceased and the pulsar companion
  started to contract to a white dwarf, the tides became inefficient
  and the rotational evolution of the companion was determined by
  conservation of angular momentum. Thus, the rotational periods scale
  inversely with the moments of inertia.

  The moment of inertia of the progenitor can be separated in that of
  the core and the envelope;
  $I_\mathrm{prog}=I_\mathrm{core}+I_\mathrm{env}$ with
  $I_\mathrm{core}=k_\mathrm{core}^2 M_\mathrm{core}^{\phantom{0}}
  R_\mathrm{core}^2$ and $I_\mathrm{env}=k_\mathrm{env}^2
  M_\mathrm{env}^{\phantom{0}} R_\mathrm{env}^2$, where $k$ is the
  gyration radius. As the progenitor fills its Roche-lobe of radius
  $R_\mathrm{L}$, we have $R_\mathrm{env}=R_\mathrm{L}$. After mass
  transfer, when the envelope has contracted to form the white dwarf,
  the white dwarf has $I_\mathrm{WD}=k_\mathrm{WD}^2
  M_\mathrm{WD}^{\phantom{0}} R_\mathrm{WD}^2$. Assuming that
  $I_\mathrm{core}\approx I_\mathrm{WD}$ and ignoring differences in
  the gyration radii, conservation of angular momentum gives
  $\Omega_\mathrm{rot}/ \Omega_\mathrm{orb}\simeq
  1+M_\mathrm{env}^{\phantom{0}}
  R_\mathrm{L}^2/M_\mathrm{WD}^{\phantom{0}} R_\mathrm{WD}^2$.  Here,
  two assumptions lead to an overestimate of the spin up; the envelope
  will be more centrally concentrated than the white dwarf, such that
  $k_\mathrm{env}<k_\mathrm{WD}$, while tidal dissipation will be
  important in the initial stages of contraction. On the other hand,
  the hot core of the progenitor will be larger than the white dwarf,
  so $I_\mathrm{core}>I_\mathrm{WD}$ (though generally this effect
  should be small, since in most cases $I_\mathrm{core}\ll
  I_\mathrm{env}$).

  For a white dwarf with a mass of 0.17--0.18\,$M_\odot$ and a radius
  of 0.042--0.058\,$R_\odot$, and for the observed mass ratio of
  $q=7.36$ and orbital period of 20\,h, the Roche-lobe radius of the
  progenitor is about $R_L=0.96\,R_\odot$. After cessation of mass
  transfer, the remaining envelope has a mass of about
  $0.01\,M_\odot$.  With these values we estimate that the rotational
  period of the white dwarf is about 20--30 times faster than the
  orbital period, so $P_\mathrm{rot}\approx1$ to 0.6\,h. In that case,
  the rotational velocity $v_\mathrm{rot}\sin i$ would be of the order
  of 50--100\,\kms. This is smaller than the 300\,\kms\ estimated
  above, and since our assumptions likely led us to overestimate the
  rotational velocity, it seems unlikely that rotational broadening
  alone could explain the discrepancy between the surface gravity
  inferred from the spectrum and that inferred from the radius
  assuming that \psr\ is a member of \gc.

  \subsection{Constraints from radio timing}\label{ssec:b.3}
  \citet{dpf+02} give two arguments for the association of \psr\ with
  \gc. The first one is that it was discovered in an observation of a
  globular cluster: the dedicated globular cluster observations with
  Parkes together cover a relatively small area compared to the whole
  sky, and the detection of a rare millisecond pulsar in this area
  suggests membership of the cluster. The problem is that the
  dedicated cluster observations are much deeper than most other
  pulsar observations, and that the number of millisecond pulsars at
  the flux level of \psr, 0.22\,mJy at 1400\,MHz, and their
  distribution on the sky, are not known.  Estimates based on
  extrapolation are uncertain. The accuracy of the estimate by D'Amico
  et al.\ of a $10^{-5}$ probability for a chance coincidence is
  therefore not clear. \citet{dpf+02} used the center beam of the
  Parkes multibeam receiver in their discovery observation. The
  diameter of that beam is about 14\arcmin\ \citep{mlc+01}, and thus
  {\em any} pulsar detected in the globular cluster survey must lie
  within 7\arcmin\ from the cluster center. We conclude that the
  argument from the probability of chance coincidence is less solid
  than the number $10^{-5}$ might suggest.

  The second argument of \citet{dpf+02} is that \psr\ has a dispersion
  measure $\mathrm{DM}=33.7$\,pc\,cm$^{-3}$, almost the same as the
  dispersion measure $\mathrm{DM}=33.3$\,pc\,cm$^{-3}$\ of the three
  pulsars in the cluster core, two of which certainly belong to the
  cluster as proven by a negative period derivative. According to the
  \citet{tc93} model, the maximum dispersion measure in the direction
  of \gc\ and \psr\ is
  $\mathrm{DM_{max}}\approx42$\,pc\,cm$^{-3}$. Since the $\mathrm{DM}$
  values of the pulsars in the core of \gc\ (and of \psr) are less
  than this, the pulsar would be almost at the distance of the
  cluster. However, the maximum to the dispersion measure arises
  because the electron layer in our Galaxy has a finite scale height
  of 0.5 to 1\,kpc, and this implies that all pulsars above the
  electron layer have the same dispersion measure in the same
  direction \citep{bv91}. Since \gc\ is at a distance of $d=4.14$\,kpc
  and a Galactic latitude of $b=-25\fdg6$, it is well above the
  electron layer, and its observed $\mathrm{DM}$ must be equated with
  the actual $\mathrm{DM_{max}}$ in that direction. (This is
  compatible with the uncertainty of about $\sqrt{4\mathrm{DM_{max}}}$
  in the model value of $\mathrm{DM_{max}}$, which gives
  13\,pc\,cm$^{-3}$ for $\mathrm{DM_{max}}\approx42$\,pc\,cm$^{-3}$;
  \citealt{nhv+97}). We conclude that the dispersion measure of \psr\
  does not prove that it is a member of the cluster, but merely that
  it is above the galactic electron layer, i.e., at a distance
  $d\ga2.4$\,kpc.

  \subsection{White dwarf models}\label{ssec:b.4}
  Finally, we cannot discard the possibility that the white-dwarf
  models themselves are uncertain. This can already be seen from the
  mass-radius relations shown in Fig.~\ref{fig:6}. These relations are
  for the observed temperature of $T_\mathrm{eff} = 10090$\,K, but,
  for a given mass of say, 0.20\,M$_\odot$, the predicted radii show a
  spread of about 0.01\,R$_\odot$.

  A part of this problem is the lack of low-mass, helium-core white
  dwarfs for which accurate parameters have been determined and which
  could be used to calibrate the evolution, cooling and atmospheric
  properties of these systems.

  \subsection{Summary and prospects}\label{ssec:b.5}
  Summarizing the results of this section, we see that, at face value,
  the systematic radial velocity and radius estimate indicates that
  \psr\ is not a member of \gc. Furthermore, we have argued that the
  similarity in dispersion measure for all all five pulsars located
  towards \gc\ does not necessarily imply that they are all at the
  same distance.

  However, our observations do not conclusively rule out the
  membership of the pulsar and the globular cluster either. All
  constraints are consistent at the $2\sigma$ level, and the
  inconsistencies of the constraints may be removed when we take into
  account that there is an allowed range in radial velocity due to the
  velocity dispersion, that there may be broadening of the Balmer
  lines in the spectrum of the white dwarf and that there are possible
  uncertainties in the white-dwarf models themselves.  As such, it is
  for future observations to decide between either possibility.  


\begin{thebibliography}{62}
    \expandafter\ifx\csname natexlab\endcsname\relax\def\natexlab#1{#1}\fi
    
  \bibitem[{{Bassa} {et~al.}(2003{\natexlab{a}}){Bassa}, {van
	Kerkwijk}, \& {Kulkarni}}]{bkk03} {Bassa}, C.~G., {van
	Kerkwijk}, M.~H., \& {Kulkarni}, S.~R. 2003{\natexlab{a}},
	\aap, 403, 1067
    
  \bibitem[{{Bassa} {et~al.}(2003{\natexlab{b}}){Bassa}, {Verbunt},
      {van Kerkwijk}, \& {Homer}}]{bvkh03} {Bassa}, C.~G., {Verbunt},
      F., {van Kerkwijk}, M.~H., \& {Homer}, L.  2003{\natexlab{b}},
      \aap, 409, L31
    
  \bibitem[{{Bessell}(1990)}]{bes90} {Bessell}, M.~S. 1990, \pasp,
    102, 1181
    
  \bibitem[{{Bessell} {et~al.}(1998){Bessell}, {Castelli}, \&
    {Plez}}]{bcp98} {Bessell}, M.~S., {Castelli}, F., \& {Plez},
    B. 1998, \aap, 333, 231
    
  \bibitem[{{Bhattacharya} \& {Verbunt}(1991)}]{bv91} {Bhattacharya},
    D. \& {Verbunt}, F. 1991, \aap, 242, 128
    
  \bibitem[{{Callanan} {et~al.}(1998){Callanan}, {Garnavich}, \&
      {Koester}}]{cgk98} {Callanan}, P.~J., {Garnavich}, P.~M., \&
    {Koester}, D. 1998, \mnras, 298, 207
    
  \bibitem[{{Carretta} {et~al.}(2000){Carretta}, {Gratton},
      {Clementini}, \& {Fusi Pecci}}]{cgcf00} {Carretta}, E., {Gratton},
    R.~G., {Clementini}, G., \& {Fusi Pecci}, F. 2000, \apj, 533, 215
    
  \bibitem[{{Cocozza} {et~al.}(2006){Cocozza}, {Ferraro}, {Possenti}, \&
      {D'Amico}}]{cfpd06} {Cocozza}, G., {Ferraro}, F.~R., {Possenti}, A.,
    \& {D'Amico}, N. 2006, \apjl, 641, L129
    
  \bibitem[{{Colpi} {et~al.}(2002){Colpi}, {Possenti}, \&
      {Gualandris}}]{cpg02} {Colpi}, M., {Possenti}, A., \& {Gualandris},
    A. 2002, \apjl, 570, L85
    
  \bibitem[{{D'Amico} {et~al.}(2001){D'Amico}, {Lyne}, {Manchester},
      {Possenti}, \& {Camilo}}]{dlm+01} {D'Amico}, N., {Lyne}, A.~G.,
    {Manchester}, R.~N., {Possenti}, A., \& {Camilo}, F. 2001, \apjl,
    548, L171
    
  \bibitem[{{D'Amico} {et~al.}(2002){D'Amico}, {Possenti}, {Fici},
      {Manchester}, {Lyne}, {Camilo}, \& {Sarkissian}}]{dpf+02} {D'Amico},
    N., {Possenti}, A., {Fici}, L., {et~al.} 2002, \apjl, 570, L89
    
  \bibitem[{{Driebe} {et~al.}(1998){Driebe}, {Sch\"onberner},
      {Bl\"ocker}, \& {Herwig}}]{dsbh98} {Driebe}, T., {Sch\"onberner},
    D., {Bl\"ocker}, T., \& {Herwig}, F. 1998, \aap, 339, 123
    
  \bibitem[{{Dubath} {et~al.}(1997){Dubath}, {Meylan}, \&
      {Mayor}}]{dmm97} {Dubath}, P., {Meylan}, G., \& {Mayor}, M. 1997,
    \aap, 324, 505
    
  \bibitem[{{Ergma} {et~al.}(1998){Ergma}, {Sarna}, \&
      {Antipova}}]{esa98} {Ergma}, E., {Sarna}, M.~J., \& {Antipova},
    J. 1998, \mnras, 300, 352
    
  \bibitem[{{Ferraro} {et~al.}(2003){Ferraro}, {Possenti}, {Sabbi}, \&
      {D'Amico}}]{fpsd03} {Ferraro}, F.~R., {Possenti}, A., {Sabbi}, E.,
    \& {D'Amico}, N. 2003, \apjl, 596, L211
    
  \bibitem[{{Filippenko}(1982)}]{fil82} {Filippenko}, A.~V. 1982, \pasp,
    94, 715
    
  \bibitem[{{Finley} {et~al.}(1997){Finley}, {Koester}, \&
      {Basri}}]{fkb97} {Finley}, D.~S., {Koester}, D., \& {Basri}, G. 1997,
    \apj, 488, 375
    
  \bibitem[{{Freire} {et~al.}(2005){Freire}, {Hessels}, {Nice},
      {Ransom}, {Lorimer}, \& {Stairs}}]{fhn+05} {Freire}, P.~C.~C.,
    {Hessels}, J.~W.~T., {Nice}, D.~J., {et~al.} 2005, \apj, 621, 959
    
  \bibitem[{{Girardi} {et~al.}(2000){Girardi}, {Bressan}, {Bertelli}, \&
      {Chiosi}}]{gbbc00} {Girardi}, L., {Bressan}, A., {Bertelli}, G., \&
    {Chiosi}, C. 2000, \aaps, 141, 371
    
  \bibitem[{{Gratton} {et~al.}(2003){Gratton}, {Bragaglia}, {Carretta},
      {Clementini}, {Desidera}, {Grundahl}, \& {Lucatello}}]{gbc+03}
    {Gratton}, R.~G., {Bragaglia}, A., {Carretta}, E., {et~al.} 2003,
    \aap, 408, 529
    
  \bibitem[{{Gratton} {et~al.}(2005){Gratton}, {Bragaglia}, {Carretta},
      {de Angeli}, {Lucatello}, {Piotto}, \& {Recio Blanco}}]{gbc+05}
    {Gratton}, R.~G., {Bragaglia}, A., {Carretta}, E., {et~al.} 2005,
    \aap, 440, 901
    
  \bibitem[{{Hamuy} {et~al.}(1994){Hamuy}, {Suntzeff}, {Heathcote},
      {Walker}, {Gigoux}, \& {Phillips}}]{hsh+94} {Hamuy}, M., {Suntzeff},
    N.~B., {Heathcote}, S.~R., {et~al.} 1994, \pasp, 106, 566
    
  \bibitem[{{Hamuy} {et~al.}(1992){Hamuy}, {Walker}, {Suntzeff},
      {Gigoux}, {Heathcote}, \& {Phillips}}]{hws+92} {Hamuy}, M.,
    {Walker}, A.~R., {Suntzeff}, N.~B., {et~al.} 1992, \pasp, 104, 533
    
  \bibitem[{{Homeier} {et~al.}(1998){Homeier}, {Koester}, {Hagen},
      {Jordan}, {Heber}, {Engels}, {Reimers}, \& {Dreizler}}]{hkh+98}
    {Homeier}, D., {Koester}, D., {Hagen}, H.-J., {et~al.} 1998, \aap,
    338, 563
    
  \bibitem[{{Horne}(1986)}]{hor86} {Horne}, K. 1986, \pasp, 98, 609
    
  \bibitem[{{Hummer} \& {Mihalas}(1988)}]{hm88} {Hummer}, D.~G. \&
    {Mihalas}, D. 1988, \apj, 331, 794
    
  \bibitem[{{Jacoby} {et~al.}(2005){Jacoby}, {Hotan}, {Bailes}, {Ord},
      \& {Kulkarni}}]{jhb+05} {Jacoby}, B.~A., {Hotan}, A., {Bailes}, M.,
    {Ord}, S., \& {Kulkarni}, S.~R.  2005, \apjl, 629, L113
    
  \bibitem[{{Joss} {et~al.}(1987){Joss}, {Rappaport}, \&
      {Lewis}}]{jrl87} {Joss}, P.~C., {Rappaport}, S., \& {Lewis}, W. 1987,
    \apj, 319, 180
    
  \bibitem[{{Koester} {et~al.}(2005){Koester}, {Napiwotzki}, {Voss},
      {Homeier}, \& {Reimers}}]{knv+05} {Koester}, D., {Napiwotzki}, R.,
    {Voss}, B., {Homeier}, D., \& {Reimers}, D.  2005, \aap, 439, 317
    
  \bibitem[{{Landolt}(1992)}]{lan92} {Landolt}, A.~U. 1992, \aj, 104,
    340
    
  \bibitem[{{Lattimer} \& {Prakash}(2004)}]{lp04} {Lattimer}, J.~M. \&
    {Prakash}, M. 2004, Science, 304, 536
    
  \bibitem[{{Le Borgne} {et~al.}(2003){Le Borgne}, {Bruzual}, {Pell{\'
	  o}}, {Lan{\c c}on}, {Rocca-Volmerange}, {Sanahuja}, {Schaerer},
      {Soubiran}, \& {V{\'{\i}}lchez-G{\' o}mez}}]{bbp+03} {Le Borgne},
    J.-F., {Bruzual}, G., {Pell{\' o}}, R., {et~al.} 2003, \aap, 402,
    433
    
  \bibitem[{{Lugger} {et~al.}(1995){Lugger}, {Cohn}, \&
      {Grindlay}}]{lcg95} {Lugger}, P.~M., {Cohn}, H.~N., \& {Grindlay},
    J.~E. 1995, \apj, 439, 191
    
  \bibitem[{{Lyne} {et~al.}(2004){Lyne}, {Burgay}, {Kramer}, {Possenti},
      {Manchester}, {Camilo}, {McLaughlin}, {Lorimer}, {D'Amico}, {Joshi},
      {Reynolds}, \& {Freire}}]{lbk+04} {Lyne}, A.~G., {Burgay}, M.,
    {Kramer}, M., {et~al.} 2004, Science, 303, 1153
    
  \bibitem[{{Manchester} {et~al.}(2001){Manchester}, {Lyne}, {Camilo},
      {Bell}, {Kaspi}, {D'Amico}, {McKay}, {Crawford}, {Stairs},
      {Possenti}, {Kramer}, \& {Sheppard}}]{mlc+01} {Manchester}, R.~N.,
    {Lyne}, A.~G., {Camilo}, F., {et~al.} 2001, \mnras, 328, 17
    
  \bibitem[{{Mihalas} {et~al.}(1988){Mihalas}, {Dappen}, \&
      {Hummer}}]{mdh88} {Mihalas}, D., {Dappen}, W., \& {Hummer},
    D.~G. 1988, \apj, 331, 815
    
  \bibitem[{{Mihalas} {et~al.}(1990){Mihalas}, {Hummer}, {Mihalas}, \&
      {Daeppen}}]{mhmd90} {Mihalas}, D., {Hummer}, D.~G., {Mihalas},
    B.~W., \& {Daeppen}, W. 1990, \apj, 350, 300
    
  \bibitem[{{Nelemans} {et~al.}(1997){Nelemans}, {Hartman}, {Verbunt},
      {Bhattacharya}, \& {Wijers}}]{nhv+97} {Nelemans}, G., {Hartman},
    J.~W., {Verbunt}, F., {Bhattacharya}, D., \& {Wijers},
    R.~A.~M.~J. 1997, \aap, 322, 489
    
  \bibitem[{{Nice} {et~al.}(2005{\natexlab{a}}){Nice}, {Splaver}, \&
      {Stairs}}]{nss05} {Nice}, D.~J., {Splaver}, E.~M., \& {Stairs},
    I.~H. 2005{\natexlab{a}}, in ASP Conference Series, Vol. 328, Binary
    Radio Pulsars, ed. F.~A. Rasio \& I.~H.  Stairs, 371
    
  \bibitem[{{Nice} {et~al.}(2005{\natexlab{b}}){Nice}, {Splaver},
      {Stairs}, {L{\"o}hmer}, {Jessner}, {Kramer}, \& {Cordes}}]{nss+05}
    {Nice}, D.~J., {Splaver}, E.~M., {Stairs}, I.~H., {et~al.}
    2005{\natexlab{b}}, \apj, 634, 1242
    
  \bibitem[{{Oke}(1990)}]{oke90} {Oke}, J.~B. 1990, \aj, 99, 1621
    
  \bibitem[{{Panei} {et~al.}(2000){Panei}, {Althaus}, \&
      {Benvenuto}}]{pab00} {Panei}, J.~A., {Althaus}, L.~G., \& {Benvenuto},
    O.~G. 2000, \aap, 353, 970
    
  \bibitem[{{Phinney} \& {Kulkarni}(1994)}]{pk94} {Phinney}, E.~S. \&
    {Kulkarni}, S.~R. 1994, \araa, 32, 591
    
  \bibitem[{{Press} {et~al.}(1992){Press}, {Teukolsky}, {Vetterling}, \&
      {Flannery}}]{ptvf92} {Press}, W.~H., {Teukolsky}, S.~A.,
    {Vetterling}, W.~T., \& {Flannery}, B.~P.  1992, {Numerical recipes:
      The art of scientific computing} (Cambridge: University Press, 2nd
    ed.)
    
  \bibitem[{{Reid}(1996)}]{rei96} {Reid}, I.~N. 1996, \aj, 111, 2000
    
  \bibitem[{{Renzini} {et~al.}(1996){Renzini}, {Bragaglia}, {Ferraro},
      {Gilmoz zi}, {Ortolani}, {Holberg}, {Liebert}, {Wesemael}, \&
      {Bohlin}}]{rbf+96} {Renzini}, A., {Bragaglia}, A., {Ferraro}, F.~R.,
    {et~al.} 1996, \apjl, 465, L23
    
  \bibitem[{{Rohrmann} {et~al.}(2002){Rohrmann}, {Serenelli}, {Althaus},
      \& {Benvenuto}}]{rsab02} {Rohrmann}, R.~D., {Serenelli}, A.~M.,
    {Althaus}, L.~G., \& {Benvenuto}, O.~G.  2002, \mnras, 335, 499
    
  \bibitem[{{Serenelli} {et~al.}(2002){Serenelli}, {Althaus},
      {Rohrmann}, \& {Benvenuto}}]{sarb02} {Serenelli}, A.~M., {Althaus},
    L.~G., {Rohrmann}, R.~D., \& {Benvenuto}, O.~G.  2002, \mnras, 337,
    1091
    
  \bibitem[{{Splaver} {et~al.}(2005){Splaver}, {Nice}, {Stairs},
      {Lommen}, \& {Backer}}]{sns+05} {Splaver}, E.~M., {Nice}, D.~J.,
    {Stairs}, I.~H., {Lommen}, A.~N., \& {Backer}, D.~C. 2005, \apj,
    620, 405
    
  \bibitem[{{Stairs}(2004)}]{sta04} {Stairs}, I.~H. 2004, Science, 304,
    547
    
  \bibitem[{{Stetson}(1987)}]{ste87} {Stetson}, P.~B. 1987, \pasp, 99,
    191
    
  \bibitem[{{Stetson}(2000)}]{ste00} {Stetson}, P.~B. 2000, \pasp, 112,
    925
    
  \bibitem[{{Tauris} \& {Savonije}(1999)}]{ts99} {Tauris}, T.~M. \&
    {Savonije}, G.~J. 1999, \aap, 350, 928
    
  \bibitem[{{Taylor} \& {Cordes}(1993)}]{tc93} {Taylor}, J.~H. \&
    {Cordes}, J.~M. 1993, \apj, 411, 674
    
  \bibitem[{{Thorsett} \& {Chakrabarty}(1999)}]{tc99} {Thorsett},
    S.~E. \& {Chakrabarty}, D. 1999, \apj, 512, 288
    
  \bibitem[{{van Kerkwijk} {et~al.}(2005){van Kerkwijk}, {Bassa},
      {Jacoby}, \& {Jonker}}]{kbjj05} {van Kerkwijk}, M.~H., {Bassa},
    C.~G., {Jacoby}, B.~A., \& {Jonker}, P.~G.  2005, in ASP Conference
    Series, Vol. 328, Binary Radio Pulsars, ed. F.~A.  Rasio \&
    I.~H. Stairs (ASP), 357
    
  \bibitem[{{van Kerkwijk} {et~al.}(1996){van Kerkwijk}, {Bergeron}, \&
      {Kulkarni}}]{kbk96} {van Kerkwijk}, M.~H., {Bergeron}, P., \&
    {Kulkarni}, S.~R. 1996, \apjl, 467, L89
    
  \bibitem[{{van Kerkwijk} \& {Kulkarni}(1995)}]{kk95} {van Kerkwijk},
    M.~H. \& {Kulkarni}, S.~R. 1995, \apjl, 454, L141
    
  \bibitem[{{van Straten} {et~al.}(2001){van Straten}, {Bailes},
      {Britton}, {Kulkarni}, {Anderson}, {Manchester}, \&
      {Sarkissian}}]{sbb+01} {van Straten}, W., {Bailes}, M., {Britton},
    M., {et~al.} 2001, \nat, 412, 158
    
  \bibitem[{{Vauclair} {et~al.}(1997){Vauclair}, {Schmidt}, {Koester},
      \& {Allard}}]{vska97} {Vauclair}, G., {Schmidt}, H., {Koester}, D.,
    \& {Allard}, N. 1997, \aap, 325, 1055
    
  \bibitem[{Verbunt(1993)}]{ver93} Verbunt, F. 1993, ARA\&A, 31, 93
    
  \bibitem[{{Wesemael} {et~al.}(1993){Wesemael}, {Greenstein},
      {Liebert}, {Lamontagne}, {Fontaine}, {Bergeron}, \&
      {Glaspey}}]{wgl+93} {Wesemael}, F., {Greenstein}, J.~L., {Liebert},
    J., {et~al.} 1993, \pasp, 105, 761
    
  \end{thebibliography}
\end{document}